\documentclass[fleqn,usenatbib]{mnras}

\usepackage{newtxtext,newtxmath}

\usepackage{fixltx2e} 
\usepackage{graphicx, amsmath, amssymb, aas_macros, natbib, tabularx}
\newcommand{\mvir}{M_{\rm{vir}}}

\newcommand{\mstar}{M_{\star}}
\newcommand{\msun}{M_{\odot}}

\newcommand{\mpc}{{\rm Mpc}}

\newcommand{\lcdm}{$\Lambda$CDM}

\newcommand{\msunyr}{\msun\,{\rm yr^{-1}}}

\newcommand{\sfr}{\dot{M}_{\star}}
\newcommand{\muv}{{M_{\rm 1500}}}
\newcommand{\hst}{\textit{HST}}
\newcommand{\jwst}{\textit{JWST}}

\newcommand{\mhalo}{M_{\rm halo}}

\newcommand{\Mgcs}{M_{\rm GCs}}

\newcommand{\mgc}{m_{\rm GC}}
\newcommand{\mgcavg}{\langle m_{\rm GC} \rangle}
\newcommand{\Mgc}{\langle M_{\rm GC}\rangle}
\newcommand{\ngcs}{N_{\rm GCs}}

\newcommand{\mprog}{M_{\rm prog}}
\newcommand{\mprogxz}{M_{\rm prog, X}}
\newcommand{\mnine}{M_{\rm prog,9.03}}
\newcommand{\mmin}{M_{\rm min}}
\newcommand{\mmp}{M_{\rm mmp}}

\newcommand{\etab}{\eta_{\rm b}}

\title[The Globular Cluster -- Dark Matter Halo Connection]
{The Globular Cluster -- Dark Matter Halo Connection
}
\author[M. Boylan-Kolchin]
{Michael Boylan-Kolchin\\
$\!$Department of Astronomy, The University of Texas at Austin,
2515 Speedway, Stop C1400, Austin, TX 78712-1205, USA; 
\href{mailto:mbk@astro.as.utexas.edu}{mbk@astro.as.utexas.edu}}

\date{Draft version, \today}

\pubyear{2017}

\begin{document}
\label{firstpage}
\pagerange{\pageref{firstpage}--\pageref{lastpage}}
\maketitle

\begin{abstract}
  I present a simple phenomenological model for the observed linear scaling of
  the stellar mass in old globular clusters (GCs) with $z=0$ halo mass in which
  the stellar mass in GCs scales linearly with \textit{progenitor} halo mass at
  $z=6$ above a minimum halo mass for GC formation. This model reproduces the
  observed $\Mgcs-\mhalo$ relation at $z=0$ and results in a prediction for the
  minimum halo mass at $z=6$ required for hosting one GC:
  $\mmin(z=6)=1.07 \times 10^9\,\msun$. Translated to $z=0$, the mean threshold
  mass is $\mhalo(z=0) \approx 2\times 10^{10}\,\msun$. I explore the
  observability of GCs in the reionization era and their contribution to cosmic
  reionization, both of which depend sensitively on the (unknown) ratio of GC
  birth mass to present-day stellar mass, $\xi$. Based on current detections of
  $z \ga 6$ objects with $M_{1500} < -17$, values of $\xi > 10$ are strongly
  disfavored; this, in turn, has potentially important implications for GC
  formation scenarios. Even for low values of $\xi$, some observed high-$z$
  galaxies may actually be GCs, complicating estimates of reionization-era
  galaxy ultraviolet luminosity functions and constraints on dark matter
  models. GCs are likely important reionization sources if $5 \la \xi \la 10$. I
  also explore predictions for the fraction of accreted versus \textit{in situ}
  GCs in the local Universe and for descendants of systems at the halo mass
  threshold of GC formation (dwarf galaxies). An appealing feature of the model
  presented here is the ability to make predictions for GC properties based
  solely on dark matter halo merger trees.
\end{abstract}


\section{Introduction} 
Globular clusters (GCs) have long challenged galaxy formation models, motivating
a variety of explanations regarding their origin and properties (e.g.,
\citealt{peebles1968, gunn1980, mccrea1982, peebles1984a, fall1985,
  rosenblatt1988, kang1990, ashman1992, murray1992, cen2001, bromm2002,
  kravtsov2005, muratov2010, elmegreen2012, kruijssen2015, kimm2016,
  popa2016}). One reason is their apparent simplicity: for many years,
metal-poor (blue) GCs were thought to be essentially simple stellar populations,
the result of a single intense star formation event. In recent years, however,
overwhelming evidence has accumulated that many or even most GCs host multiple
populations, indicating at least two generations of star formation. In
particular, the existence of internal spreads in the light-element abundances of
GCs is incompatible with the original picture of each globular as a stellar
population with a single age and metallicity (see, e.g., \citealt{gratton2012}
and \citealt{renzini2015} for recent overviews).

One of the most enigmatic aspects of GCs is their formation and its relationship
to dark matter halos. Many models of GC formation posit that GCs form at the
centers of dark matter halos during early phases of galaxy formation
\citep{peebles1984a, rosenblatt1988, moore2006}, yet there is no dynamical
evidence for dark matter in globular clusters \citep{moore1996a, conroy2011a}.
Nevertheless, intriguing hints of connections between globular clusters and dark
matter halos exist. The strongest of these is the relationship between the total
mass in globular clusters within a dark matter halo, $\Mgcs(z=0)$, and the
halo's mass, $\mhalo$: \citet{hudson2014} and \citet{harris2017} recently showed
that $\Mgcs=\eta\,\mhalo$, with $\eta=3-4\times 10^{-5}$ using the
\citet{harris2013} database of GCs (see \citealt{blakeslee1997, spitler2009,
  georgiev2010, harris2013} for similar results using smaller samples of halos
and clusters and \citealt{kravtsov2005} for related results based on
cosmological simulations). This observation is somewhat surprising, at first
glance: metal-poor ($[{\rm Fe/H}] \la -1.1]$) globular clusters are thought to
form at high redshift ($z \ga 5$; \citealt{brodie2006, vandenberg2013,
  forbes2015}), meaning that if their formation is somehow connected with halo
mass, the quantity to correlate with is $\mhalo$ at high redshift, not at
$z=0$. Furthermore, halos grow differentially as a function of halo mass,
meaning that a linear correlation at $z=0$ would be a non-linear correlation at
high redshifts.

In this paper, I propose a simple model for explaining the $\Mgcs-\mhalo(z=0)$
relation -- all halos above a minimum mass at high redshift are capable of
forming a (metal-poor) globular cluster, with total stellar mass in globular
clusters at formation proportional to dark matter halo mass at that epoch -- and
explore some consequences for the low-redshift and high-redshift Universe.

\section{Assumptions}
\label{sec:assumptions}
Masses of individual globular clusters will be denoted with a lower-case $m$,
while the mass of a globular cluster system will be denoted with an upper-case
$M$. Dependence on redshift will be explicitly noted where relevant. Average
values will be indicated with angle brackets $\langle \ldots \rangle$. For example,
$\langle m(z=6) \rangle$ means the average mass of a globular cluster at $z=6$,
while $\langle \Mgcs| \mhalo (z=6) \rangle$ indicates the average total stellar
mass in globular clusters in a halo of mass $\mhalo$ at $z=6$. I will assume the
following in my subsequent analysis:
\begin{itemize}
\item The present-day mass function of globular clusters in a given galaxy is
  lognormal, with $\log_{10}$-mean
  $\langle \log_{10}(m/\msun) \rangle \equiv \mu \approx 5.2$ and
  $\log_{10}$-dispersion $\sigma \approx 0.5$ (e.g., \citealt{harris2017}):
  \begin{equation}
    \label{eq:3}
    \frac{dN_{\rm GCs}}{d\log_{10} m}=\frac{1}{\sqrt{2\,\pi\,\sigma^2}}\,
    \exp\left[-\frac{(\log_{10}m-\mu)^{2}}{2\sigma^2}\right]\,.
  \end{equation}
  Note that this makes the mean mass of a globular cluster
  $\langle m(z=0) \rangle = 10^{\mu+\log(10)\,\sigma^2/2}$. However, common
  practice is to define $\langle m(z=0) \rangle \equiv \Mgcs/N_{\rm GCs}$ with
  $N_{\rm GCs}$ defined to be twice the number of GCs with stellar masses
  greater than $10^{\mu}$ under the assumption of a constant stellar
    mass to light ratio for all GCs; I will assume
  $\langle m(z=0) \rangle=2.5\times 10^5\,\msun$ (see \citealt{harris2017}).
\item The present-day stellar mass of a globular cluster is smaller than its
  stellar mass at birth by a factor $\xi$:
  \begin{equation}
    \label{eq:5}
\frac{\langle \mgc({\rm birth}) \rangle} {\langle \mgc(z=0) \rangle }
    \equiv \xi\,.
  \end{equation}
  Implicit in this assumption is that the initial mass function of globular
  clusters can be approximated by a log-normal distribution. However, with the
  definition of $\langle \mgc(z=0) \rangle$ adopted above, this is unimportant. Models 
  explaining light-element abundance spreads in GCs often require $\xi \sim
  10-100$ (e.g., \citealt{bekki2007, dercole2008, conroy2012a}, though see
  \citealt{bastian2015} for arguments against large values of $\xi$); I will
  generally leave $\xi$ as an unknown parameter and investigate implications for
  $\xi \approx 1-3$ and $\xi \sim 10-100$ separately.
\item The total mass in present-day globular clusters within a dark matter halo
  is a constant fraction of the mass of the dark matter halo for
  $10^{10}\la\mhalo/\msun\la 10^{15}$:
\begin{equation}
  \label{eq:6}
 \langle \Mgcs | \mhalo(z=0) \rangle =\eta\,\mhalo(z=0)\,
\end{equation}
with $\eta=(3-4)\times 10^{-5}$ (\citealt{hudson2014, harris2015, harris2017}; see also
\citealt{spitler2009, georgiev2010, harris2013}). 
\item The total mass in present-day \textit{metal-poor} globular clusters
  within a dark matter halo is also a constant fraction of the mass of the dark
  matter halo:
\begin{equation}
  \label{eq:mgcs_bgcs}
 \langle \Mgcs | \mhalo(z=0) \rangle =\etab\,\mhalo(z=0)\,
\end{equation}
with $\etab \approx (2-2.5) \times 10^{-5}$ \citep{harris2015, harris2017}.
\item The epoch of formation for metal-poor globular clusters coincides with the
  era of reionization and ends at $z=6$.
\item Dark matter halo merger histories are well-approximated by the extended
  Press-Schechter \citeyearpar{press1974} model using the Parkinson et al.
  \citeyearpar{parkinson2008} algorithm.\footnote{I am very grateful to Yu Lu
    for providing code for generating merger trees.} I generate merger trees in
  the \citet{planck2015} cosmology for halos with
  $10 \le \log_{10}(\mhalo(z=0)/\msun \le 14$ with spacing of 0.33 dex; each
  $z=0$ mass is sampled with 100 trees and a minimum resolved halo mass of
  $10^7\,\msun$. For a subset of $z=0$ masses, I generate 1000 trees to obtain
  improved statistics.
\end{itemize}

\section{A Simple Model}
\label{sec:model}
\subsection{Forming Globular Clusters at High Redshifts}
The primary assumption of this work is simple: the mass in globular cluster
stars in a halo at $z \approx 6$ is linearly related to the mass of dark matter
halo at that time and that there is some minimum mass $\mmin$ below which
globular cluster formation is strongly suppressed (or is
impossible).

The average stellar mass in globular clusters within a given halo in this model
is
\begin{equation}
  \label{eq:13}
  \langle \Mgcs | \mhalo(z=6) \rangle= \langle \mgc(z=6)\rangle\,
  \frac{\mhalo}{\mmin}\,
  f\left(\frac{\mhalo}{\mmin}\right)\,, 
\end{equation}
where $f(x)$ is a suppression function (that could be 0 for $\mhalo(z=6) <
\mmin$ and 1 otherwise or, more likely, is a smooth function with an exponential
suppression below $\mmin$).
The relationship between the (average)
total mass in globular cluster stars and halo mass at $z=6$ is then given by
\begin{equation}
  \label{eq:14}
  \langle \Mgcs | \mhalo(z=6) \rangle = \langle \mgc(z=6)\rangle\, 
\frac{\mhalo}{\mmin}
\end{equation}
for $\mhalo/\mmin \gg 1$. In other words, the total stellar mass in
  globular cluster stars in a given halo at the end of the epoch of globular
  cluster formation is proportional to ratio the halo mass at that time and
  $\mmin$. The average \textit{number} of globular clusters hosted by a halo of
mass $\mhalo$ at $z=6$ follows directly from Eq.~\ref{eq:13} and the
  definition of $\ngcs$ given in the first bullet point of
  Section~\ref{sec:assumptions}: 
\begin{equation}
  \label{eq:12}
  \langle \ngcs | \mhalo(z=6) \rangle =\frac{\mhalo}{\mmin}\, 
  f\left(\frac{\mhalo}{\mmin}\right)\,.
\end{equation}

\begin{figure}
 \centering
 \includegraphics[width=\columnwidth]{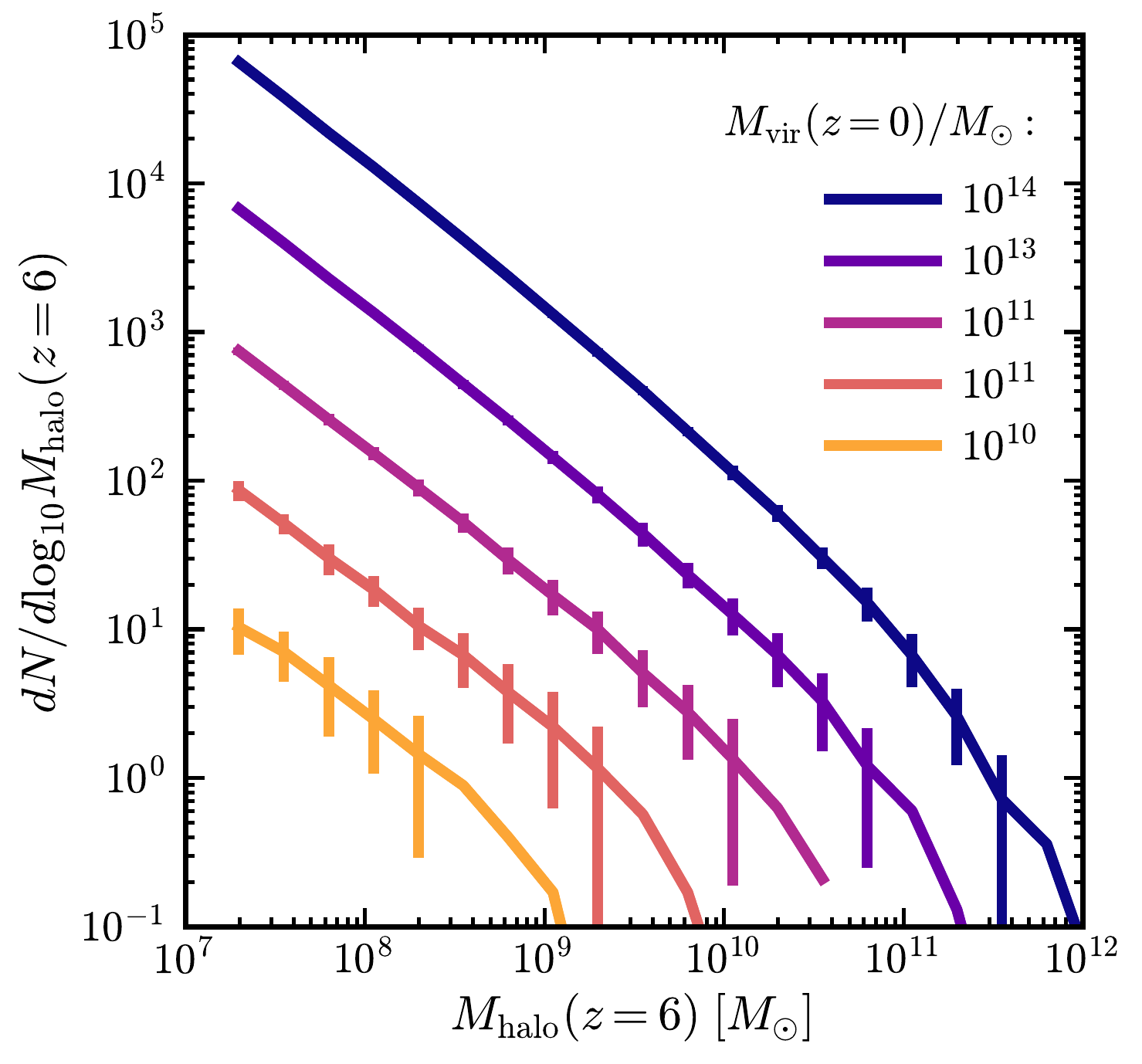}
 \caption{The average conditional (progenitor) dark matter halo mass function at $z=6$ for a
   variety of $z=0$ halo masses. Note the similarity in shapes and that the
   normalization at fixed $\mhalo(z=6)$ scales linearly with $\mvir(z=0)$.
 \label{fig:mfunc}
}
\end{figure}

\begin{figure*}
 \centering
 \includegraphics[width=\columnwidth]{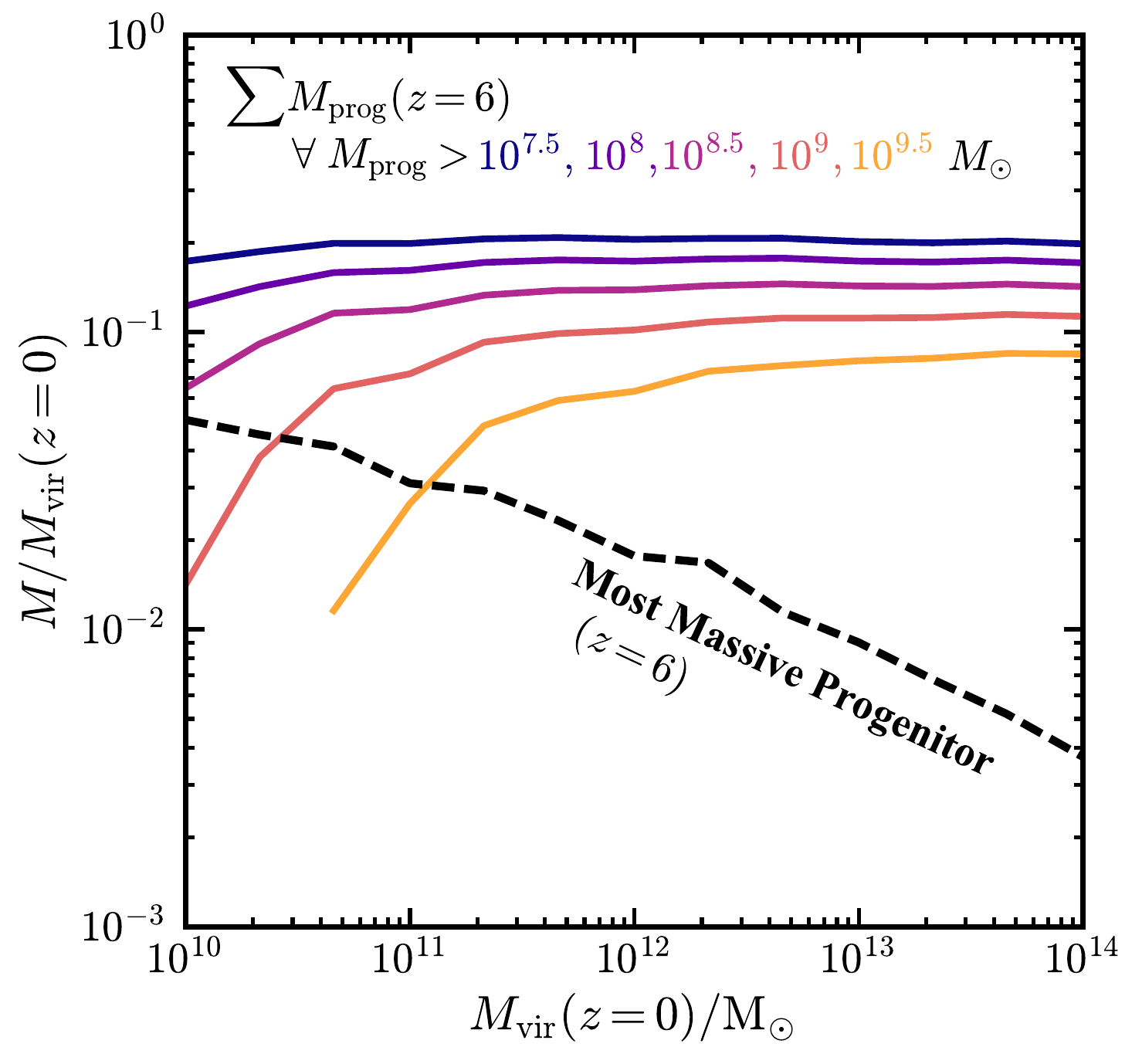}
 \includegraphics[width=\columnwidth]{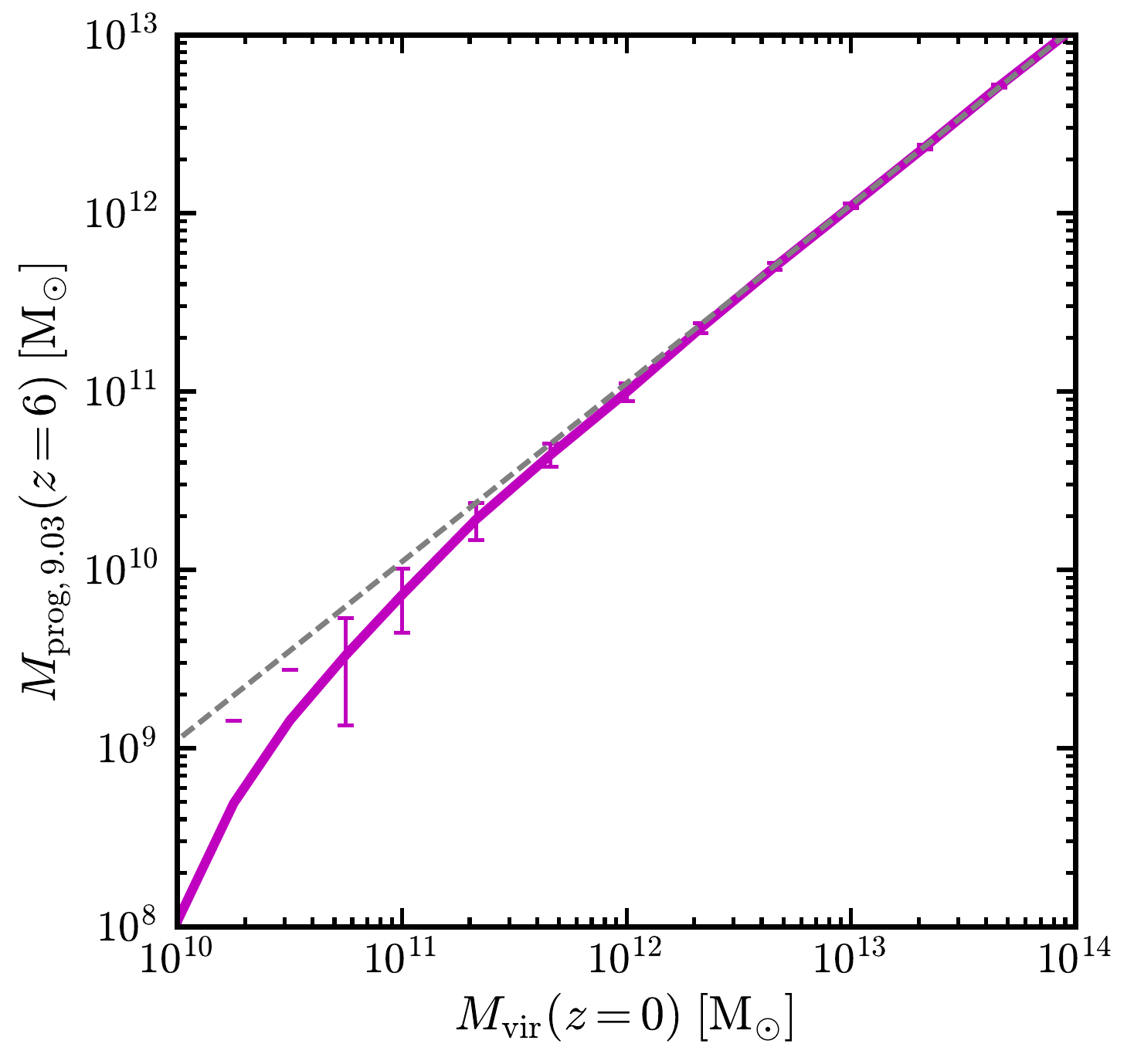}
 \caption{\textit{Left:} A comparison of the halo mass contained within progenitors
   above given thresholds at $z=6$ ($\mprogxz$; solid lines) with 
   the average mass of a halo's main progenitor at $z=6$ (dashed line), 
both relative to the
   $z=0$ halo mass. While the ratio of most massive progenitor to $z=0$ mass
   declines with $\mhalo(z=0)$, $\mprogxz/\mhalo(z=0)$ is essentially constant
   over the entire range. Furthermore, the absolute value of $\mprogxz/\mhalo(z=0)$
   only differs by a factor of $\approx 2$ across 2 decades in threshold
   mass. Models directly relating $\mprogxz$ to $\Mgcs$ naturally reproduce the
   observed correlation between $\Mgcs$ and $\mhalo(z=0)$. \textit{Right:} The
   sum of $\mhalo$ over all progenitors with 
   $\mhalo(z=6) > \mmin=1.07\times 10^9\,\msun$ as a function of
   $\mhalo(z=0)$. The dashed line shows $M_{\rm prog,9.03} \propto
   \mhalo(z=0)$. For 
   $\mhalo \gg \mmin$, there is a one-to-one correlation. At lower masses, the
   halo-to-halo scatter becomes relevant, and below
   $\mvir(z=0) \approx 3\times 10^{10}\,\msun$, many $z=0$ halos do not have
   even 1 progenitor above $\mmin$. See Eq.~\eqref{eq:19} for a fitting
   formula for $\mnine$ given $\mhalo(z=0)$.
 \label{fig:mprog6_vs_mvir_z0}
}
\end{figure*}

\subsection{Evolving to $z=0$}
At $z=0$, halos are composed of many progenitors of mass
$\mprog(z) < \mhalo(z=0)$. If (metal-poor) globular cluster formation is limited
solely to the reionization era, then relations imprinted on the dark matter halo
population at $z=6$ can be propagated to $z=0$ simply by understanding the dark
matter assembly histories of $z=0$ halos. In particular, the total mass in
globular clusters stars as a function of halo mass is simply the sum of the mass
in GCs in progenitors above $\mmin$ (ignoring, for now, GC
disruption):
\begin{equation}
  \label{eq:15}
  \Mgcs(z=0)=\sum_{i}M_{\rm GCs}^i(z=6)\,,
\end{equation}
where the sum is performed over all progenitors having $\mhalo^{i}(z=6) >
\mmin$. The average mass in globular clusters in a halo as a
function of $\mhalo(z=0)$ will be 
\begin{flalign}
  \label{eq:1}
  \langle \Mgcs|\mhalo(z=0) \rangle &= \sum_i \langle \Mgcs | \mhalo^i(z=6) \rangle \\
&=\langle \mgc(z=0) \rangle \sum_i \frac{\mhalo^i}{\mmin}\,f\left(\frac{\mhalo^i}{\mmin}\right)\\
&\equiv\langle \mgc(z=0) \rangle\frac{\mprogxz(z=6)}{\mmin}\,.\label{eq:mgcs}
\end{flalign}
In other words, the total mass of globular cluster stars in a halo of mass
$\mhalo$ at $z=0$ is directly related to the sum of the masses of all its
progenitor halos with $M(z=6) > \mmin$. In the nomenclature I adopt, $\mprogxz$
is the sum in progenitors at $z=6$ with $\log_{10}(M/\msun) > X$.

The minimum mass of a halo hosting a GC at high $z$ is defined by
$\mmin=10^X$. Equating Eq.~\ref{eq:6} and Eq.~\ref{eq:mgcs} yields
\begin{equation}
  \label{eq:10}
  \mmin=\frac{\mgcavg}{\etab}\,\frac{\mprogxz}{\mhalo(z=0)}= 10^{10}\msun\,
  \frac{\mprogxz}{\mhalo(z=0)}\,, 
\end{equation}
where the second equality follows from my default assumptions of
$\mgc=2.5\times 10^5\,\msun$ and $\etab=2.5\times 10^{-5}$. Note that
Eq.~\ref{eq:10} is sensitive to the combination $\mgcavg/\etab$ and that there
is implicit dependence on $\mmin$ in $\mprogxz$. 

The full results of the merger trees give $\mprogxz(\mmin)$: at $z=6$, the
dependence of $\mprogxz$ on $\mmin$ can be approximated by
  \begin{equation}
    \label{eq:16}
  \frac{\mprogxz}{\mhalo(z=0)}=0.11\,(\mmin/10^9\,\msun)^{-0.184}
  \end{equation}
  to better than 5\% accuracy over the range $10^{7.5} < \mmin/\msun <
  10^{9.5}$.
I therefore obtain 
\begin{equation}
  \label{eq:17}
\mmin(z=6)\approx 1.07\times 10^9\,\msun   
\end{equation}
(i.e., $X=9.03$).  As is shown in Figure~\ref{fig:mprog6_vs_mvir_z0}, such a
model actually does give a linear correlation with $\mvir(z=0)$, essentially
irrespective of the mass threshold ($\mmin$) chosen. This is a natural
  outcome of cosmological models with CDM-like power spectra, as it is a feature
  of the high-redshift conditional mass function of $z=0$ halos (which can be
  calculated using extended Press-Schechter theory;
  e.g., \citealt{lacey1993, van-den-bosch2002}; see
  Fig.~\ref{fig:mfunc}).\\[-0.2cm]

\textit{A model in which globular clusters populate
all dark matter halos above $\mmin(z=6) \approx 10^9\,\msun$ in direct
proportion to $\mhalo(z=6)$ naturally reproduces the observed
$\mhalo-\Mgcs$ relation at $z=0$}.\\[-0.2cm]

Eq.~\ref{eq:10} can be split into dependence on dark matter
properties [through $\mprogxz/\mhalo(z=0)$] and baryonic properties (via
$\mgcavg/\etab$). Any redshift dependence enters solely through the relationship
between the mass in collapsed progenitors above a given threshold at redshift
$z$ ($\mprogxz$). In this work, I assume that the epoch of blue globular cluster
formation ends at $z=6$, and therefore this is the relevant redshift. It is
straightforward to quantify the redshift dependence of Eq.~\ref{eq:10},
however: I find
\begin{equation}
  \label{eq:18}
  \log_{10}(\mmin(z)/\msun)=-0.165\,z+10.04\,.
\end{equation}

\begin{figure}
 \centering
 \includegraphics[width=\columnwidth]{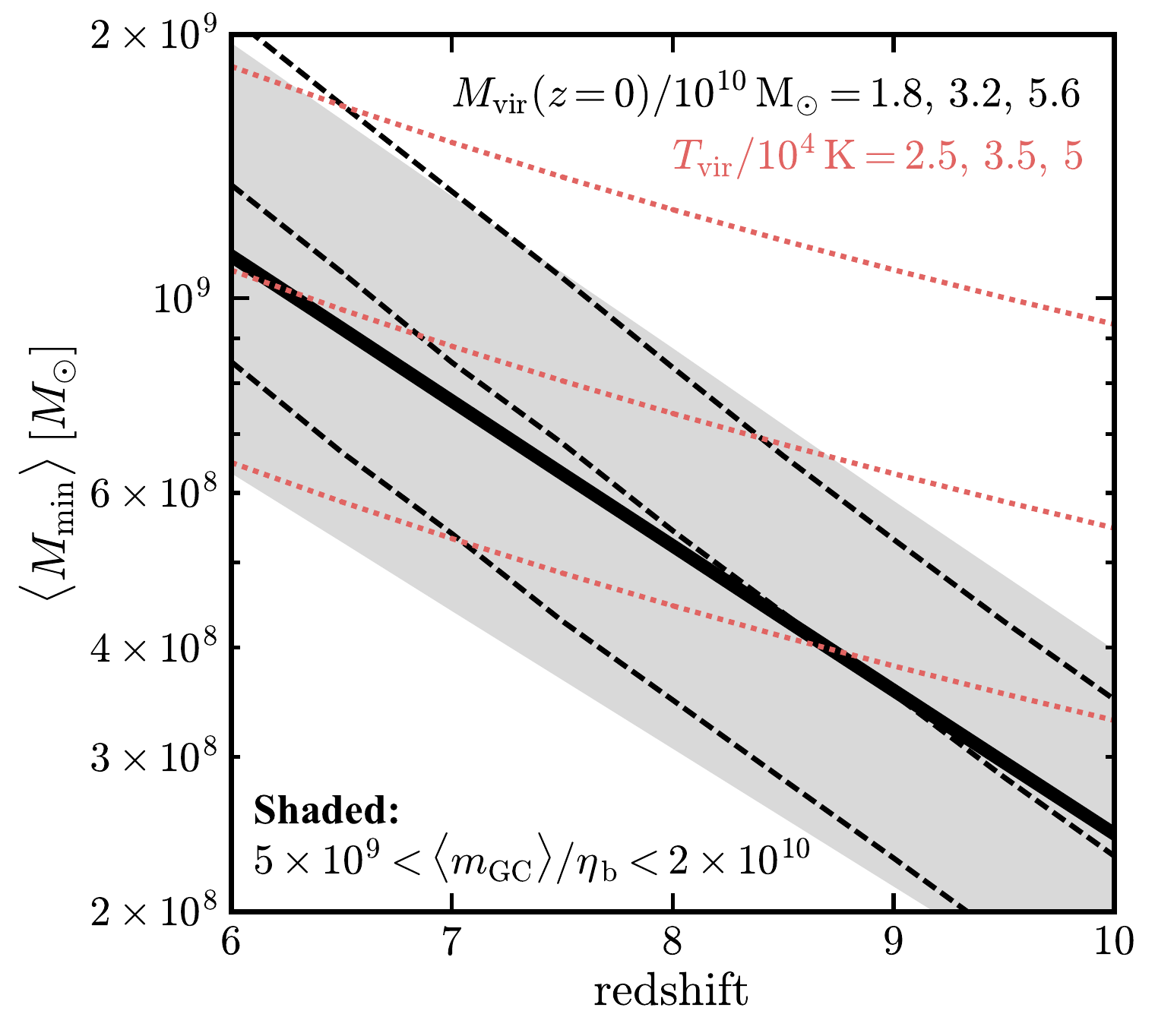}
 \caption{The dependence of $\mmin$ on redshift. Thick solid line: $\mmin(z)$,
   as obtained from solving Eq.~\ref{eq:10}. Shaded region: same, but allowing
   for a factor of 2 variation (both high and low) in $\mgcavg/\etab$. Dashed
   black lines: the average mass assembly histories for halos with
   $\mvir(z=0)/10^{10}\,\msun=1.8, \,3.2,\,5.6\,\msun$. Dotted orange lines: halos
   having virial temperatures of $T_{\rm vir}/10^{4}\,{\rm K}=2.4, \,3.5,\,
   5$. The evolution of $\mmin(z)$ is closely tracked by the evolution of the
   most massive progenitor of a typical halo with $\mvir(z=0) \approx 2.5\times
   10^{10}\,\msun$. 
 \label{fig:mminz}
}
\end{figure}

Figure~\ref{fig:mminz} shows how $\mmin$ depends on the inferred epoch of
globular cluster formation. The redshift evolution of $\mmin$ is can be
well-approximated by the redshift evolution of the most massive progenitor
($\mmp$) of
halos having $\mvir(z=0) \approx 2.5\times 10^{10}\,\msun$. Independent of the
redshift we chose to assign to GC formation, we will end up with the same
relationship between $\mhalo(z=0)$ and the mass of the globular cluster system.
The gray band in the figure shows the results of varying $\mgcavg/\etab$ by a
factor of $\pm 2$ from the fiducial value of $10^{10}\,\msun$.

The model laid out in this section is extremely simple: the total mass in
globular clusters is directly proportional to halo mass at high redshift, with a
minimum halo mass $\mmin$ capable of hosting 1 globular cluster of
$\mmin \approx 10^9\,\msun$ at $z=6$. Halos for which $\mmp(z) \gg \mmin(z)$
will be insensitive to this cut-off. Halos with $\mmp(z) \approx \mmin(z)$ will
be stochastically populated with globular clusters. Although this model is
purely phenomenological rather than rooted in the detailed physics of globular
cluster formation, its simplicity means that I can make several predictions
about globular clusters and galaxy formation. These predictions and their
implications are explored in subsequent sections.

\vspace{1cm}

\section{Implications and Predictions at High Redshifts}
\label{sec:highz}
\subsection{Prospects for Observing High-$z$ Globular Clusters}
\label{subsec:observability}
\begin{figure*}
 \centering
 \includegraphics[width=\columnwidth]{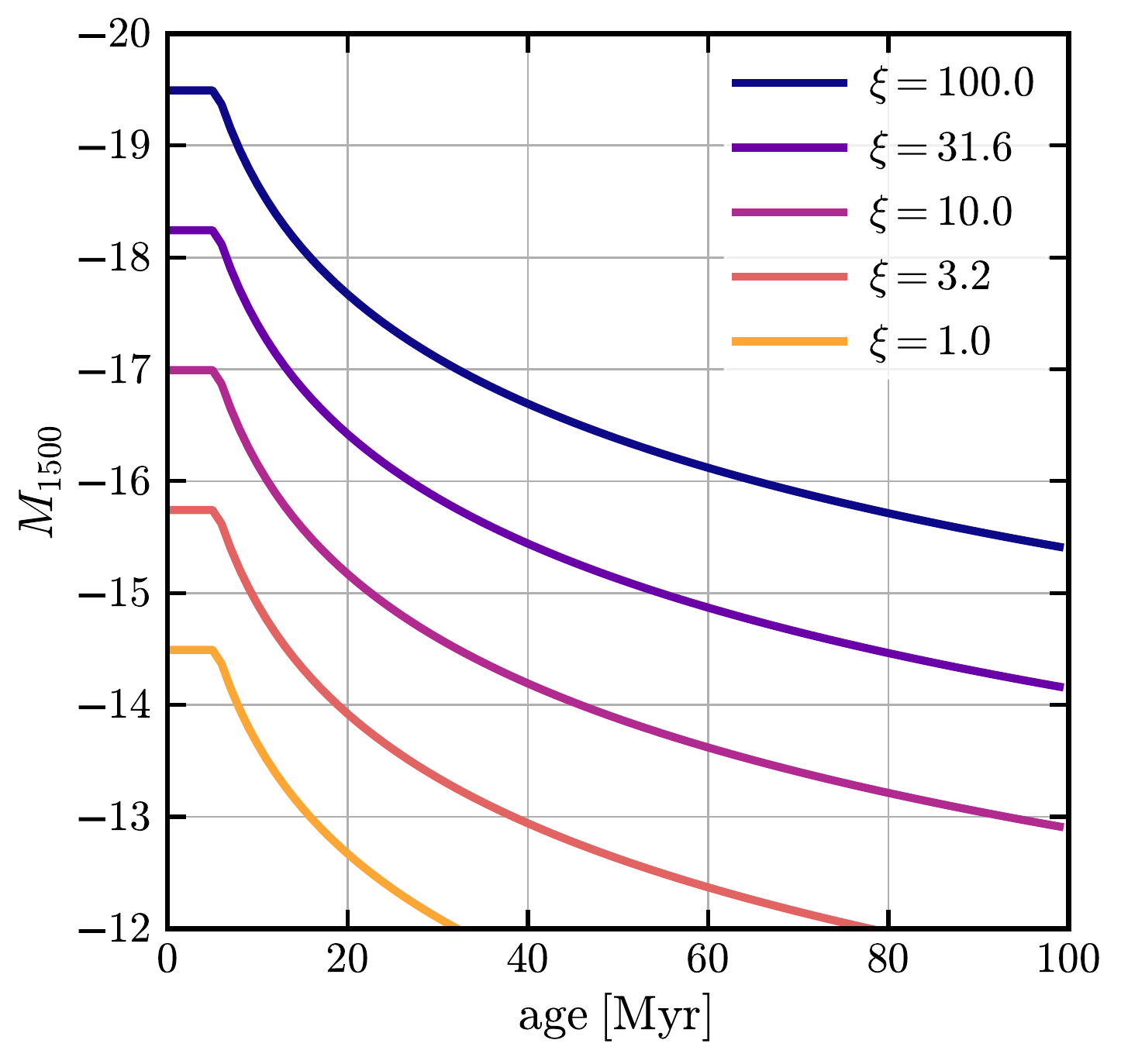}
 \includegraphics[width=\columnwidth]{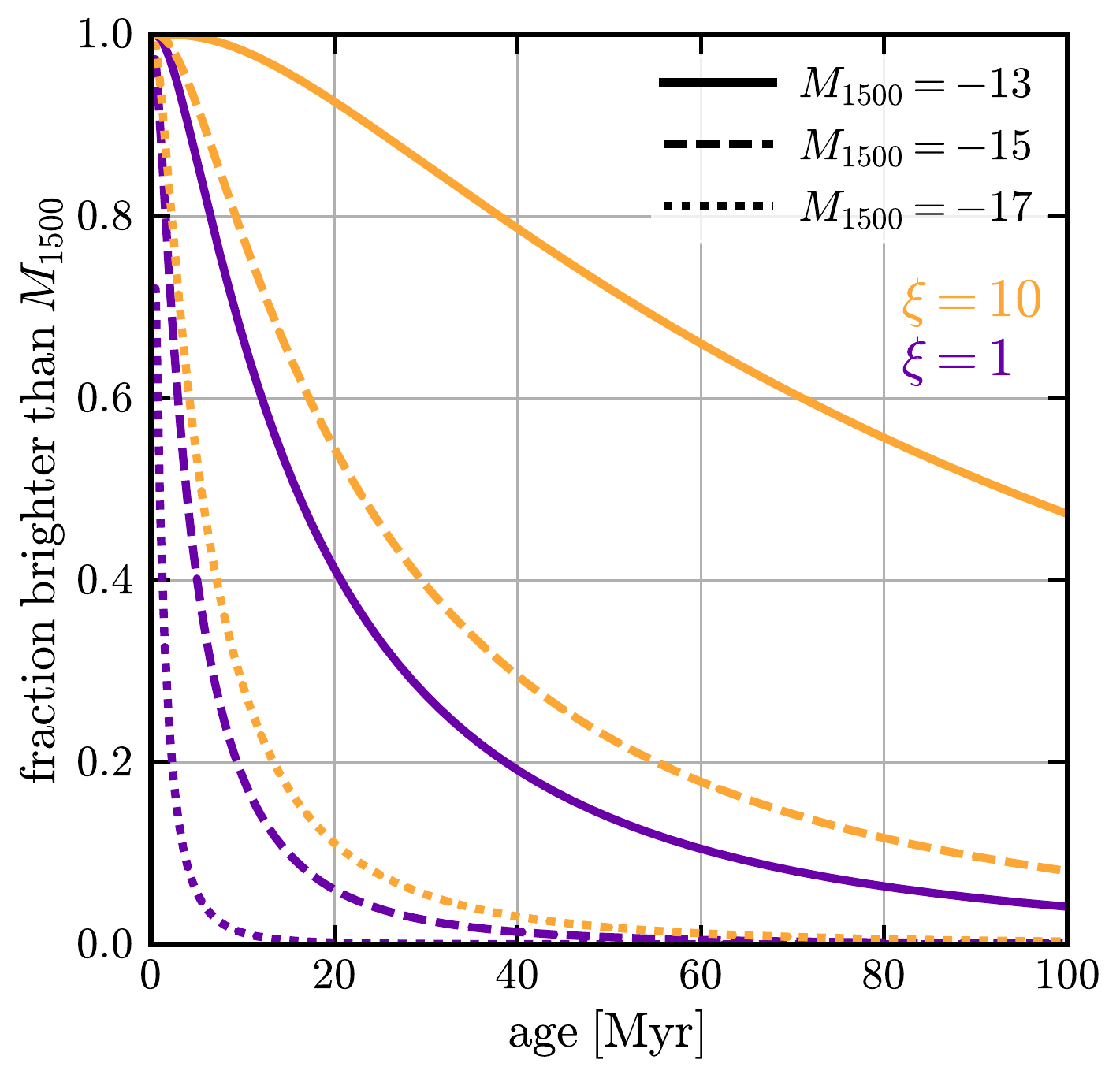}
 \caption{\textit{Left:} the absolute UV magnitude of a globular cluster with birth mass
   $\mstar=\xi\times\, 2.5\times 10^5\,\msun$ as a function of time after
   birth. $L_{1500}$ has a plateau for 5.5 Myr, then falls off as
   $t^{-1.3}$. \textit{Right:} the fraction of globular clusters (with GC mass
   function given by Eq.~\eqref{eq:3}) brighter than various $M_{1500}$
   thresholds as a function of time after completion of GC star formation for
   $\xi=1$ (purple) and $\xi=10$ (orange). This quantity is $f(<\muv|\Delta t)$, which
   enters Eq.~\eqref{eq:observability}. The effective duty cycle of the GC mass
   function (i.e., what fraction of GCs can be observed at a given time) is
   given by $f(<\muv|\Delta t) \, \Delta t/t_{\rm form}$, where $t_{\rm
     form}$ is the period over which all GCs form.
 \label{fig:observability}
}
\end{figure*}

The \textit{ansatz} of this paper -- that the average mass in metal-poor GCs in
a $z=0$ dark matter halo is directly related to the total dark matter mass in
its progenitors above a fixed threshold in the reionization era -- provides the
means to compute the comoving number density of globular clusters: this is
simply the comoving mass density of halos with $\mhalo(z=6) \ge \mmin$ divided
by $\mmin$:
\begin{equation}
  \label{eq:reion}
  n_{\rm GCs}=\frac{1}{\mmin}\int_{\mmin}^\infty dM\,\mhalo\,\frac{dn_{\rm halo}}{d\mhalo}\,.
\end{equation}
At $z=6$, $\mmin=1.07\times10^9 \,\msun$, implying a comoving cumulative number
density of $n_{\rm GCs} = 2.2\,\mpc^{-3}$ for blue globular clusters (using the
\citealt{sheth2001} mass function).  

With this number, we can investigate the observability of high-redshift galaxy
formation as traced by globular clusters. Assuming a \citet{kroupa2001} stellar
initial mass function, $[{\rm Fe/H}]=-2$, either Padova \citep{marigo2008,
  girardi2010} or MIST \citep{dotter2016, choi2016} stellar evolution models
(see below) and a 5 Myr duration for star
formation (with constant $\sfr$), a globular cluster will have an
  absolute rest-frame magnitude at 1500~\AA ($M_{1500}$) that can be
approximated by
\begin{flalign}
M_{1500}(t)&=-16.75-2.5\,\log_{10}\left(\frac{\mstar}{2\times10^6\,\msun}\,g(t)\right)
\label{eq:m1500}\\
g(t)&=\begin{cases}
1\; &{\rm if } \;t<5.5\,{\rm Myr}\\
\left(\frac{t}{5.5\,{\rm Myr}}\right)^{-1.3} & {\rm if} \;5.5\le t/{\rm Myr} \la 300\\
\end{cases}
\end{flalign} 
This result is based on calculations using the Flexible Stellar Population
Synthesis models of \citealt{conroy2009a, conroy2010a}, including nebular
emission \citep{byler2017}, and is fairly insensitive to the choice of stellar
models: the rms deviations from the fit are 0.10 mag and 0.16 mag and the
maximal deviation from the fit is 0.32 mag and 0.37 mag for Padova and MIST
models, respectively, for $t < 300\,{\rm Myr}$. I caution against using
Eq.~\ref{eq:m1500} for ages beyond 300 Myr, as the errors become substantial. In
the future, it will be important to consider variations in stellar models that
include, e.g., binaries or fast-rotating massive stars, as these models predict
a different peak value and evolution of $\muv$ with
time.\footnote{\#starsAreStillInteresting.}

The number of GCs in the redshift range $(z_1,z_2)$ is 
\begin{equation}
  \label{eq:gcs_time}
  N(z_1, z_2)=\int_{z_1}^{z_2} n(z) \,dV\,.
\end{equation}
If GCs form with a uniform distribution in time over the interval $(z_1, z_2)$,
then Eq.~\ref{eq:gcs_time} becomes
\begin{flalign}
  \label{eq:gcs_time_constant}
  N(z_1, z_2)&=\frac{n_{\rm GCs}}{t(z_1,z_2)} \int_{z_1}^{z_2} t(z_1, z)
  \frac{dV}{dz}{dz}\,\\
& \equiv  n_{\rm GCs} V(z_1, z_2) \, \zeta(z_1,z_2)\,,
\end{flalign}
where $V(z_1, z_2)$ and $t(z_1, z_2)$ are the comoving volume and cosmic time,
respectively, between redshifts $z_1$ and $z_2$; $\zeta(z_1,z_2)$ encapsulates
the effect of a changing number of GCs with $z$, i.e., how much $N(z_1,z_2)$
differs from the naive estimate of $n_{\rm GC}\,V(z_1,z_2)$ owing to the time
evolution of $n_{\rm GC}(z)$. Unless otherwise noted, I assume that GCs form
with a uniform distribution in time over $10 \ge z \ge 6$ ($t(z_1,z_2)=457$
Myr). The correction factor $\zeta$ is non-negligible, as it has a maximum value
of $0.5$ (obtained in the limit $z_1 \approx z_2$). I find
$\zeta(z_1=10,\,z_2=6)=0.44$.

The number of observable GCs in a field of angular size $\Omega$ is then
approximately 
\begin{equation}
  \label{eq:observability}
 N(<\muv)=N(z_1, z_2)\left( \frac{\Omega}{\rm 4\pi\,sr} \right)\,\frac{\Delta 
  t}{t(z_1, z_2)} \,f({<\muv|\Delta t})\,, 
\end{equation}
where $\Delta t$ is the time period for which a GC exceeds the given $\muv$
threshold and $f(<\muv|\Delta t)$ is the fraction of globular clusters brighter
than $\muv$ for a period of at least $\Delta t$ using
Eqns.~\eqref{eq:3}~and~\ref{eq:m1500}. $N(<\muv)$ is therefore a trade-off
between the observability period $\Delta t$ (which is longer for more massive
GCs) and the fraction of globular clusters brighter than $\muv$ for a time
period of at least $\Delta t$ (which is a decreasing function of $\Delta
t$). Note that for a lognormal distribution of GC masses, as assumed here,
$\sim 4.8\%$ of GCs will have masses larger than 10 times the median (turnover)
mass. For $\xi=1$, this means 5\% of GCs will reach $\muv < -17$. If $\xi=10$,
then 5\% will reach $\muv = -19.5$ (for approximately 5 Myr) and 50\% of GCs
will reach $\muv = -17$!

\begin{table}
\setlength{\tabcolsep}{13pt}
\caption{\textit{Observability of globular clusters at high redshifts.} The
  entries in the table give the number of globular clusters brighter than the
  absolute magnitude listed in column 1 for various values of $\xi$ (the
  average value of the ratio of birth mass to $z=0$ mass). The numbers quoted
  here assume $\Omega=5\,{\rm arcmin^2}$, comparable to the size of ACS on
  \hst. For \jwst, the numbers below should be multiplied by 2 ($\Omega=9.7\,{\rm
    arcmin}^2$).  At $z=6$, $M_{1500}=-15$ corresponds to $m_{1500}=31.7$; this
  is approximately the reach of a \jwst\ ultra-deep survey (200 hr). 
}
  \label{tab:table1}
  \begin{tabular}{@{   }c | r r r r r@{   }} 
    \hline
    \hline
      & \multicolumn{5}{l}{Number of Detectable Globular Clusters in $5\,{\rm arcmin^2}$} \\
      & \multicolumn{5}{l}{$(6 \le z \le 10)$}\\
    \hline
    $M_{1500}$ & $\xi=1$ & 3.16 & 10 & 31.6 & 100\\
    \hline
-17	& 20	& 48	& 116	& 282	& 684\\
-16	& 40	& 97	& 236	& 573	& 1389\\
-15	& 82	& 198	& 480	& 1163	& 2435\\
-14	& 166	& 402	& 974	& 2173	& 3314\\
-13	& 337	& 816	& 1906	& 3129	& 3796\\
    \hline
  \end{tabular}
\end{table}

The expected number of GCs detectable for various $M_{1500}$ thresholds are
listed in Table~\ref{tab:table1} for $\Omega=5\,{\rm arcmin}^2$, roughly
comparable to the field of view of
ACS\footnote{\href{http://www.stsci.edu/hst/acs/}
  {http://www.stsci.edu/hst/acs/}} on the \textit{Hubble Space Telescope}
(\hst). The numbers in Table~\ref{tab:table1} are not very sensitive to the
assumed redshift range of globular cluster formation (so long as the lower limit
is $z \approx 5-6$): for example, the quoted numbers should be multiplied by a
factor of 1.09 if $6 \le z \le 7$ or 0.87 if $5 \le z \le 15$. $N_{\rm GCs}$
scales linearly with $\Omega$, so for a field the size of
NIRCam\footnote{\href{https://jwst.stsci.edu/instrumentation/nircam}
  {https://jwst.stsci.edu/instrumentation/nircam}} on the \textit{James Webb
  Space Telescope} (\jwst) -- $9.7\,{\rm arcmin}^2$ -- the numbers are
approximately double those listed in the table.

In fact, many of the faintest gravitationally-lensed sources observed at
$z\sim 6-8$ are consistent with small sizes \citep{bouwens2017, vanzella2017},
perhaps indicating that globular clusters are the easiest objects to detect at
high redshift. At a minimum, a \jwst\ survey reaching $M_{1500}=-15$ ($m=31.7$)
should observe $\approx 160$ forming GCs per field. If $\xi>1$, the number will
be larger: for example, if $\xi=10$, then $N_{\rm GCs} \approx 1000$ per
field\footnote{As this paper was in the final stages of preparation,
  \citet[hereafter R17]{renzini2017} independently investigated the
  observability of GCs with \jwst. While some of the considerations in R17 are
  similar to those presented here, the approach is fairly different and
  complementary to that of this paper.}. Of these, essentially all are expected
to be in separate dark matter halos -- on average, $\la 1$ halo in this volume will
be massive enough to host $>1$ detectable GC at a time. The number of GCs will
therefore provide a probe of the halo mass function at high redshifts. It is
important to note that, in this context, GCs are not really distinct from
galaxies; rather, it is the large quantity of stars formed in a short time in a
globular cluster that makes a globular the easiest part of a galaxy to see. An
important distinction between GCs and ``normal'' star formation is that the
standard conversion from UV flux to star formation rate \citep{kennicutt1998,
  madau2014} assumes extended star formation of $t_{\rm sf} > 10^{7}\,{\rm yr}$,
which is not true for GCs. In the model under discussion, GCs are forming in
dark matter halos, but likely not at their centers (thereby accounting for their
observed lack of dark matter).

\hst\ surveys of the UDF have revealed approximately 150 galaxies at $z \ga 6$
with $M_{1500} \la -17$ \citep{bouwens2015, finkelstein2015}. The numbers in
Table~\ref{tab:table1} immediately disfavor models with $\xi \ga 15$: if
globular cluster birth masses were $\ga 15$ times their present-day masses, the
HUDF would have many more high-$z$ detections compared to what is actually
observed. Even $\xi \approx 10$ implies that $\ga 70\%$ of detections at
$z \ga 6$ are globular clusters rather than ``normal'' star formation in
high-$z$ galaxies. \textit{The results presented here therefore strongly suggest
  that the birth masses of globular clusters exceed their present-day stellar
  masses by less than a factor of ten.} The number counts and size distribution
of $\muv \ga -16$ objects at $z \approx 6-10$ in \jwst\ blank fields will
provide a stringent test of whether or not metal-poor GCs are much less massive
than they were at birth.

Based on the arguments above, the stellar mass completeness of UV-selected
galaxy samples at high redshift will be strongly influenced by globular clusters
for $\muv \ga -16$. One globular cluster that forms $\sim 10^6\,\msun$ of stars
in a nearly instantaneous burst ($t \approx 5\,{\rm Myr}$) with
$\sfr \approx 0.2 \,\msunyr$ and then fades passively will
move from $\muv \approx -16.5$ immediately following the burst to
$\muv \approx -14$ within 40 Myr. Note that a galaxy that forms $10^{6}\,\msun$
of stars at a constant rate over a (current) Hubble time will have a star
formation rate of $10^{-4}\,\msunyr$, which would correspond to
$\muv \approx -7.5$. Even if all of the $10^6$ stars formed over a (relatively)
short period of 1 Gyr ($\sfr \approx 10^{-3} \,\msunyr$),
which is even shorter than duration inferred for the Draco dwarf spheroidal
galaxy, it would only result in $\muv \approx -10$ while the galaxy is actively
forming stars. These considerations are relevant when considering the importance
of galaxies versus GCs as very faint sources (below \hst\ detection thresholds)
when constructing UV luminosity functions and interpreting them in the context
of reionization.

\subsection{Connections with Cosmic Reionization} 
\label{subsec:reion}
Using the average mass of a GC, we can also compute the stellar mass density in
GCs; this is $\rho_{\rm GCs}=\mgcavg\,n_{\rm GCs}=\xi\times 5.5\times
10^{5}\,\msun\,\mpc^{-3}$ at $z=6$. This
number can be related to the importance of GCs for reionization (e.g.,
\citealt{ricotti2002, schaerer2011, katz2013}). The star formation rate
density required to maintain an ionized IGM is given by
\begin{equation}
  \label{eq:11}
  \dot{\rho}_{\star}=1.2\times10^{-3}\left(\frac{C}{f_{\rm esc}}\right)
  \left(\frac{1+z}{8}\right)^3\,\msunyr\,\mpc^{-3}\,.
\end{equation}
(see eq.~27 of \citealt{madau1999}; $C$ is the clumping factor of the
intergalactic medium and $f_{\rm esc}$ is the escape fraction of ionizing
photons). If we assume that metal-poor GCs form over the epoch $z=10$ to $z=6$,
integrating equation~\ref{eq:11} gives a requisite stellar mass to maintain
reionization over that period of
$\sim 7.2\times 10^5\,\msun\,\mpc^{-3}\,(C/f_{\rm esc})$. GCs therefore provide a
fraction $0.76\,\xi\,f_{\rm esc}\,C^{-1}$ of the ionizing flux needed to
maintain reionization.

Fiducial models often assume $C=3$ and $f_{\rm esc} \approx 0.2$ (e.g.,
\citealt{robertson2015, mcquinn2016}), with uncertainties in $f_{\rm esc}$
dominating those in $C$ (e.g., \citealt{ma2015}). GCs may have even higher
escape fractions of $f_{\rm esc} \ga 0.5$ \citep{ricotti2002}, in which case,
GCs are likely major contributors to reionization irrespective of
$\xi$. Similarly, if $\xi \approx 10$, GCs are important reionization sources so
long as $f_{\rm esc} \ga 0.1$ (and potentially dominant if $f_{\rm esc} \ga 0.2$
for GCs). Even if $\xi \approx 5$, GCs may provide the bulk of ionizing
photons\footnote{GCs are also likely sites for the formation of X-ray binaries,
  which may also contribute to reionization \citep{mirabel2011}.}
at early times if $f_{\rm esc} \ga 0.3$. If both $\xi$ and $f_{\rm esc}$ are
small, GCs do not significantly contribute to reionization , though they still
may play an important role in their immediate environments. Understanding the
relationship between the stellar mass at birth and at present is therefore a
pressing question for \textit{cosmology} as well as for star formation.

In passing, I note that $\mvir(z=6) \approx 10^9$ -- the minimum mass of a halo
hosting one dark matter halo as derived above -- corresponds to
$\muv \approx -13$ in abundance-matching models \citep{kuhlen2012a,
  boylan-kolchin2014, boylan-kolchin2015, boylan-kolchin2016}; this is
approximately the UV luminosity at which \citet{boylan-kolchin2014,
  boylan-kolchin2015} and \citet{weisz2017} argue there should be a break in the
high-$z$ galaxy luminosity function based on the stellar fossil record in the
Local Group. The connection between the mass required to host, on average, 1 GC
and the turn-over halo mass for the high-$z$ UVLF is no more than circumstantial
at this point, but it is not difficult to imagine a scenario in which halos
above some critical mass have multiple channels for star formation (e.g.,
``normal'' star formation and GC star formation) while those below the mass
scale do not achieve the high gas densities required for GC formation. This
issue will be addressed in a future paper.

\subsection{Constraining the Dark Matter Power Spectrum}
\begin{figure}
 \centering
 \includegraphics[width=\columnwidth]{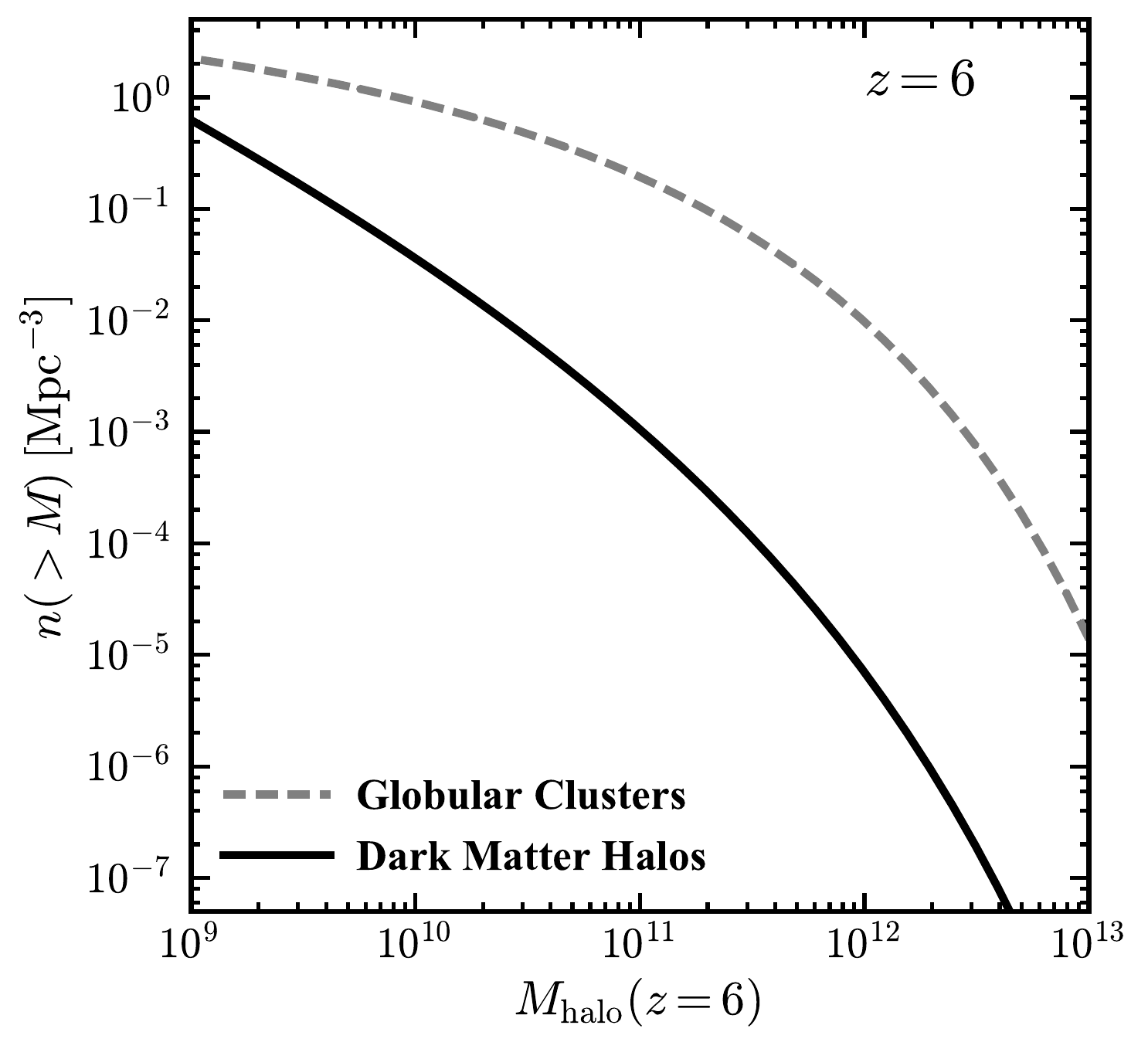}
 \caption{The cumulative number density of dark matter halos (solid black line)
   and globular clusters (dashed gray line) as a function of halo mass at $z=6$
   according to the model presented in 
   this paper. A single source detected with implied cumulative number density of $1
   \,\mpc^{-3}$ could either be a galaxy in a $7\times 10^8\,\msun$ halo or a GC
   in a $10^{10}\,\msun$ halo. The possibility of detecting globular clusters at
   high $z$ therefore 
   complicates the ability to use UV luminosity functions to constrain dark
   matter models, as the basic assumptions of abundance matching break
   down. 
   \label{fig:num_densities}
 }
\end{figure}

Various authors have pointed out that high-redshift galaxy luminosity functions
can serve as strong constraints on any cut-off in the dark matter power spectrum
at wavenumbers corresponding to halo masses of $\mvir \approx 10^{8-9}\,\msun$
\citep{schultz2014, bozek2015, menci2017}.  The basic idea is straightforward:
by assuming galaxies populate halos with a monotonic mapping between $\muv$ and
$\mhalo$, the existence of faint galaxies necessarily implies a population of
low-mass halos. Current results using this technique appear to rule out much of
the parameter space in which Warm Dark Matter models produce appreciably
different structure than Cold Dark Matter models \citep{schultz2014, menci2017};
similarly, models of ``fuzzy dark matter'' \citep{hu2000, hui2017} that
naturally produce kpc-scale cores in dwarf galaxies at low-$z$ may not have
enough structure at high-$z$ to produce the abundance of observed galaxies
\citep{bozek2015}.

If, however, GCs are significant contributors to the UV luminosity function --
as the results in this paper suggest -- then the true constraints are likely
significantly weaker. The detection of a single galaxy with an implied
cumulative number density of $1\,\mpc^{-3}$, which would rule out models that
suppress power on scales $\la 10^{9}\,\msun$ in standard abundance matching
models, can easily be attributed to a GC in a much more massive halo that is
caught $\sim 20\,{\rm Myr}$ after its peak star formation. As is shown in
Figure~\ref{fig:num_densities}, $\ngcs(>\mhalo) \approx 1\,\mpc^{-3}$ is
attained at a halo mass that is an order of magnitude greater than the
halo mass corresponding to $n(>\mhalo) \approx 1\,\mpc^{-3}$
($\mhalo \sim 10^{10}\,\msun$ versus $\mhalo \approx 10^9\,\msun$).

\subsection{Efficiency of Globular Cluster Star Formation}
\label{subsec:efficiency}
The linear relationship between the present-day mass in GCs and
the sum of $z\sim6$ dark matter halo mass in halo progenitors indicates that
GCs form at a constant efficiency in the high-redshift
universe. With a universal baryon fraction of
$f_{\rm b}=\Omega_{\rm b}/\Omega_{\rm m} \approx 0.156$, and assuming that
globular clusters had stellar masses at formation of $\xi\,\mgc(z=0)$, I obtain
a baryon conversion efficiency of $\xi\,\Mgc/(f_{\rm b}\,\mmin)$. If $\xi$ is
large ($\xi \approx 10-20$), then the baryon conversion efficiency into old
globular cluster stars is also large ($\approx 1.4-2.8\%$), meaning $\sim 2\%$
of baryons in dark matter halos more massive than $\mmin$ at $z \approx 6$ are
converted into stars in blue globular clusters. If $\xi \approx 1-2$, then
closer to $0.1\%$ of baryons in these dark matter halos are converted into stars
in blue GCs.

\section{Implications and Predictions at Low Redshift}
\label{sec:lowz}

\subsection{Globular clusters in low-mass dark matter halos}
\label{subsec:low-mass}

\begin{figure}
 \centering
 \includegraphics[width=\columnwidth]{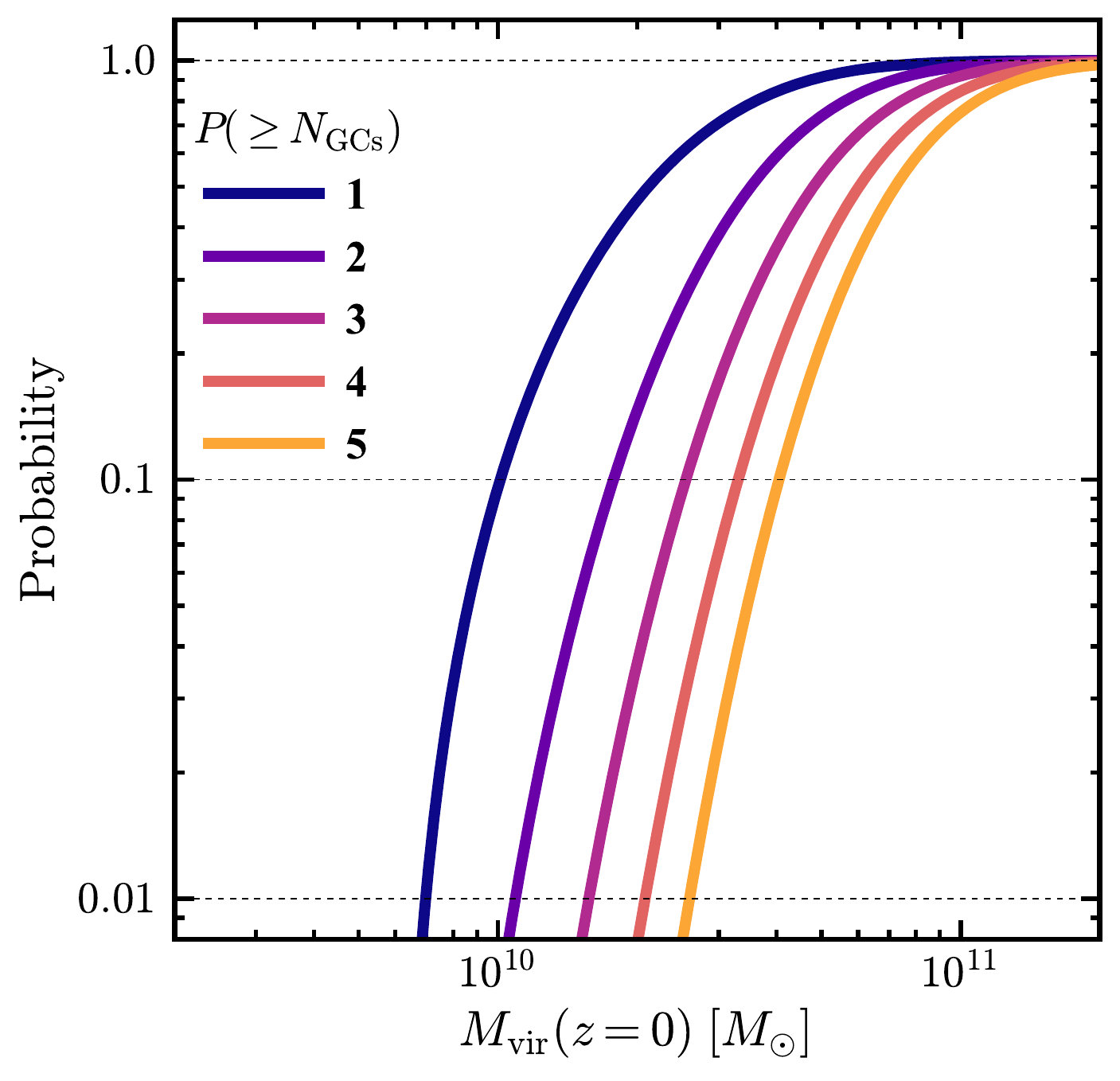}
 \caption{Probability distributions for hosting at least $N$ globular clusters
   as a function of halo mass at $z=0$ according to the model
   presented in Section~\ref{subsec:low-mass}. The mass at which $P(\ge 1\,{\rm
     GC}) = 0.5$ is $2.1\times 10^{10}\,\msun$. At lower masses, the probability
   of hosting a globular cluster is strongly suppressed in this model.
 \label{fig:neg_binom}
}
\end{figure}

A simple model for globular clusters in dark matter halos at $z=0$ is that the
mean mass in globular cluster stars as a function of halo mass is
$\mgcavg\,\mnine/\mmin$ (with $\mmin=10^{9.03}\,\msun$; see
Eq.~\ref{eq:17}). The right panel of Figure~\ref{fig:mprog6_vs_mvir_z0} shows
the dependence of $\mnine$ on $\mhalo(z=0)$; a good fit is given by
\begin{equation}
  \label{eq:19}
\mnine=0.11\,\mhalo\,\left[1-\left( \frac{\mhalo}{A\times 
      10^{9.03}\,\msun}\right)^{-b} \right]^{1/b}\,,
\end{equation}
with $A=6$ and $b=0.575$.

Observationally, the scatter in the mass in globular clusters at fixed halo
mass is approximately constant with $\sigma \la 0.28$ dex. A plausible model for
such a relation is the negative binomial distribution, which can be parametrized
by a mean value and a (constant) intrinsic scatter (see
\citealt{boylan-kolchin2010}). In the limit that the intrinsic scatter goes to
zero, the negative binomial distribution converges to a Poisson distribution. I
therefore assume that the mass in GCs is given by $\mgcavg\,\ngcs$ and that
$\ngcs$ is described by a negative binomial distribution with a mean value of
\begin{equation}
  \label{eq:20}
  \langle N \rangle = \frac{\mnine}{1.07\times 10^9\,\msun}\,
\end{equation}
and an intrinsic scatter of 40\%. This toy model reproduces the observed
$\Mgcs-\mhalo$ relationship and its observed $0.28\,{\rm dex}$ scatter
(\citealt{harris2017}, assuming a scatter of $0.2\,{\rm dex}$ in the stellar
mass-halo mass relationship; \citealt{hudson2014}).

Figure~\ref{fig:neg_binom} shows how the probability for hosting at least $N$
globular clusters varies as a function of halo mass at $z=0$ according to this
model. At $\mvir(z=0)=2.1\times10^{10}\,\msun$, there is a 50\% chance of
hosting at least one blue GC. This probability drops precipitously at lower
masses: 1 in 10 halos with $\mvir(z=0) =1.01\times 10^{10}\,\msun$ will host at
least one blue GC, while fewer than 1 in 100 halos with
$\mvir =7 \times 10^9\,\msun$ will host a blue GC. These numbers are useful in
the context of considering possible dwarf galaxy hosts of GCs near the Milky Way
\citep{zaritsky2016} and the halo masses of dwarf galaxies known to host blue
GCs, e.g., Eridanus II \citep{koposov2015, crnojevic2016} and Pegasus
\citep{cole2017}. More sophisticated (and perhaps more physically plausible)
models of the occupation number of GCs as a function of $z=0$ halo mass might
include the effects of environment on $P(\ngcs|\mhalo)$; for example, patchy
reionization might contribute to scatter in formation redshifts for globular
clusters \citep{spitler2012}.

\subsection{\textit{In-Situ} versus accreted globular clusters}
\label{subsec:insitu}
\begin{figure}
 \centering
 \includegraphics[width=\columnwidth]{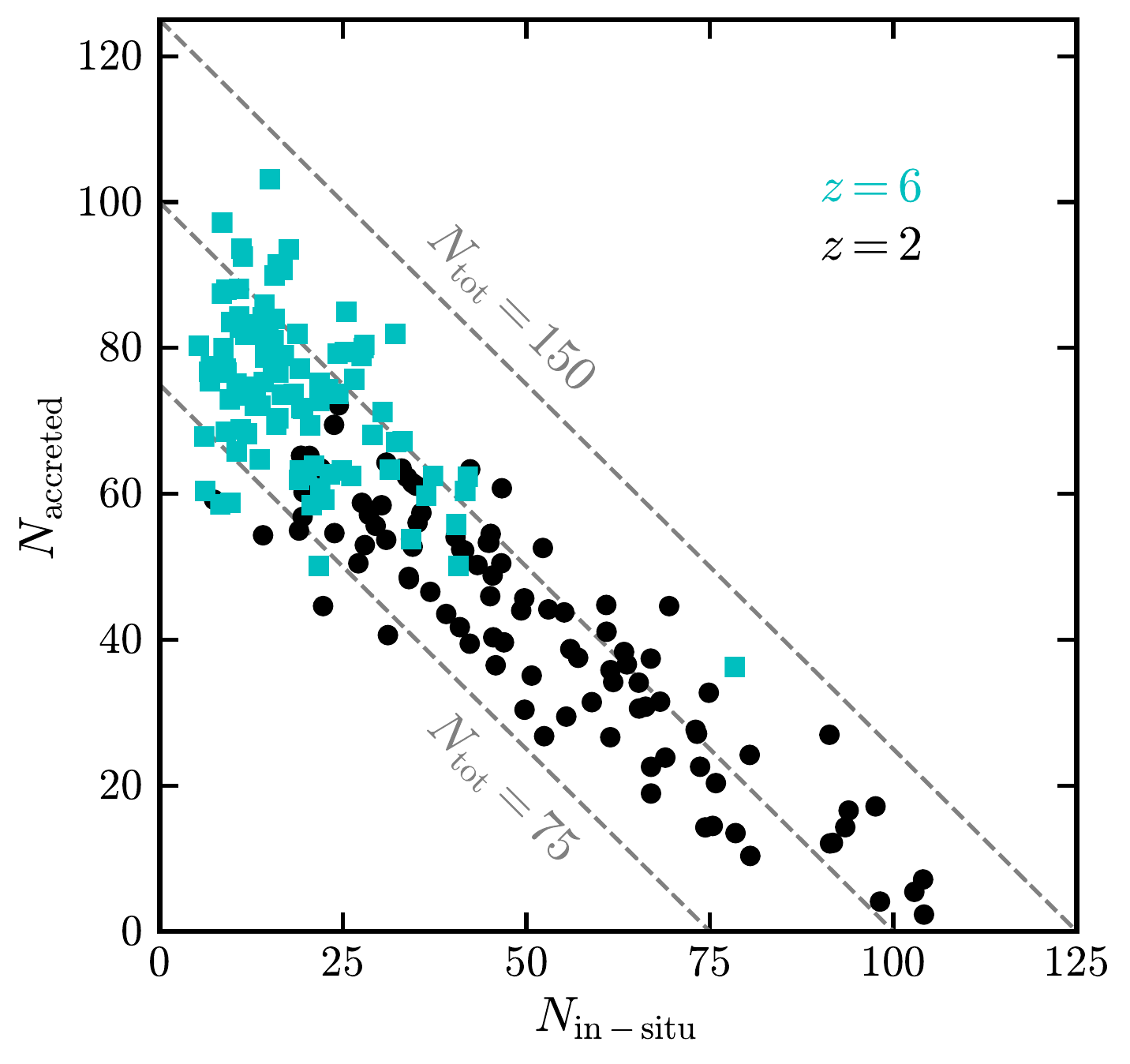}
 \caption{The number of in-situ globular clusters (those in the main progenitor)
   versus accreted GCs at $z=6$ (cyan) and $z=2$ (black) for Milky-Way-mass
   halos at $z=0$. At $z=6$, the majority of GCs are not in the main progenitor;
   by $z=2$, more than half of the realizations contain at least $50\%$ of the
   $z=0$ blue GCs in the main progenitor. 
 \label{fig:insitu}
}
\end{figure}

My model for globular cluster formation also enables a straightforward
calculation of the fraction of globular clusters formed in-situ by a given
redshift (i.e., that formed by redshift $z$ in the main progenitor of a
$z=0$ halo) versus those that were formed in a separate halo and later
accreted. The average fraction of in-situ clusters for a given halo mass at
$z=6$ is simply $\langle \mmp \rangle/\mnine$. The average fraction of in-situ
clusters determined using any later redshift $z$ requires calculating the
fraction of mass in progenitors at $z=6$ above $\mmin$ that ends up in the main
progenitor at $z$; this is straightforward, given a merger tree.

Figure~\ref{fig:insitu} shows the predicted number of in-situ versus accreted
blue globular clusters for 100 realizations of Milky-Way-mass halos
($\mhalo(z=0)=10^{12}\,\msun$). The cyan points show results at $z=6$, while the
black points show results for $z=2$. At early times, most of the globular
clusters are \textit{not} associated with the Milky Way's main progenitor; in
other words, most of the blue globular clusters of the present-day Milky Way
were not formed in-situ. I find that, on average, 16.6\% of GCs found in the
Milky Way today were formed in its main progenitor by $z=6$, with a 90\%
confidence interval of 8-40\%. The upper end of this range agrees with the
somewhat different models of \citet{katz2014}. The typical accreted GC came in
as part of a relatively low-mass progenitor: the average dark matter halo
contributing to the accreted population hosts 2 GCs, and there are typically
$\sim 30$ such progenitors. If we consider $z=2$, there is a much broader
range of possibilities, from the vast majority of the Milky Way's blue GCs
already residing in the main progenitor to the majority coming from accretion
subsequent to $z=2$. In general, halos with earlier assembly histories will have
a larger fraction of their globular clusters in place (in situ) at a given
redshift. Since the Milky Way appears to have a relatively quiescent recent
merger history, it may tend toward the region of Figure~\ref{fig:insitu} that is
dominated by in-situ globulars at $z=2$.

Another interesting regime is the low-$\mhalo$ end of the $\mhalo-\Mgcs$
relation (see also \citealt{zaritsky2016}). This is particularly true in the
context the Fornax dwarf spheroidal galaxy, which has 5 globular clusters (4 of
which are metal-poor), all of which reside within 2 kpc of the galaxy's center
(in projection; \citealt{mackey2003, de-boer2016}). The very existence of these
globulars is somewhat surprising on multiple levels: the implied specific
frequency of GCs in Fornax is very high (e.g., \citealt{georgiev2009}), and
furthermore, dynamical friction should have caused them to spiral to the center
of the galaxy if they have been present in the galaxy for a Hubble time
\citep{tremaine1976a, oh2000}. One explanation of their survival is that Fornax
has a cored dark matter distribution \citep{goerdt2006}; alternatively, perhaps
the globular clusters in Fornax have been accreted on a cosmologically recent
timescale (e.g., \citealt{cole2012}), in which case the dynamical friction
arguments would be weakened substantially.

In the context of the model presented here, it is extremely unlikely that
\textit{all four} of Fornax's metal-poor were accreted recently. At
$\mhalo \approx 3\times10^{10}\,\msun$, 2\% of merger tree realizations have at
least 4 GCs. Of these 2\%, approximately 1/4 have at least 2 accreted (post
$z=6$) GCs, but all of these have at least 2 in-situ clusters as well. Only
0.2\% of the realizations have even one GC accreted after $z=2$.  This is a
consequence of the relatively sharp cut-off in the abundance of globulars in
galaxies below the mass of Fornax.

\citet{harris2017} noted that the $\Mgcs-\mhalo$ relation predicts a halo mass
of $\sim1.5\times 10^{10}\,\msun$ for Fornax, while many dynamical estimates of
its halo mass based on stellar kinematics indicate a much lower mass
($0.5-1\times 10^9\,\msun$; \citealt{penarrubia2008, kuhlen2010,
  boylan-kolchin2012}). They interpreted this as a possible break down in the
$\Mgcs-\mhalo$ relation for low-luminosity galaxies. However, the $\Mgcs-\mhalo$
prediction for Fornax's halo mass is roughly in line with abundance matching
expectations \citep{garrison-kimmel2017, read2017}, perhaps indicating that the
constancy of $\etab$ persists even to the dwarf galaxy regime. It is important
to note that dynamical estimates of Fornax's mass are based on stellar
kinematics, which only extend to $\sim 1$ kpc from its center. In \lcdm, Fornax
should be embedded in a much more extended (and massive) dark matter halo,
reconciling its apparently low halo mass with the $\Mgcs-\mhalo$
relationship. The high specific frequency of globular clusters in Fornax and
other dwarf galaxies is somewhat of a red herring in this model: it is an
outcome of the low \textit{overall} star formation efficiency (per halo mass) in
dwarf galaxies combined with the constant globular cluster star formation
efficiency.

\section{Summary}
The possible association of (metal-poor) globular clusters with dark matter
halos has a venerable history, starting with
\citet{peebles1984a}. \citet{rosenblatt1988} suggested that associating GCs with
$2.8\,\sigma$ density fluctuations in CDM models naturally reproduces, to first
order, various properties of the Milky Way's GC system. \citet{moore2006} argued
that the spatial distribution and kinematics of both GCs and dwarf galaxies in
the Milky Way is indicative of formation in peaks of at least $2.5\,\sigma$ in
the density distribution (see also \citealt{boley2009, corbett-moran2014}). They
also noted that ``the mass fraction in peaks of a given $\sigma$ is independent
of the final halo mass'' above some minimum host mass.

In this paper, I present a phenomenological model that is rooted in this
lineage. The model is simple: I assume that blue globular clusters form in halos
above some minimum mass $\mmin$ by $z=6$ and that the mass in globular cluster
stars formed is directly proportional to the dark matter halo mass at
formation. GCs need not form at the centers of the halos (and likely do not,
given existing constraints on the dark matter content of GCs); rather, they may
be the subcomponents of a halo's baryonic content that have fragmented out of
compressed gas and would coexist with ``normal'' star formation in halos at high
redshifts. In such a model, the linear correlation between $\Mgcs$ and $\mhalo$
at $z=0$ is naturally reproduced, which is a consequence of the mass assembly of
CDM halos and the CDM power spectrum. This model also predicts the minimum mass
of a dark matter halo capable of hosting one GC during the epoch of GC
formation; for the parameters adopted here, it is
$\mmin(z=6)=1.07\times 10^9\,\msun$ (with mild redshift dependence). Using this
basic model, I explore a number of implications and predictions at both high and
low redshifts:

\begin{itemize}
\item \textbf{High-$z$ observability}: I estimate the number density of old GCs
  to be $\approx 2\,\mpc^{-3}$ using the formalism of
  Sections~\ref{sec:assumptions}-\ref{sec:model}. If the epoch of formation for
  metal-poor GCs spans $10>z>6$, then high-$z$ globular cluster formation should
  be visible directly, as GCs are UV-bright for a period that depends only on
  the total mass in stars formed (i.e., on $\xi$, the ratio of GC birth mass to
  present-day mass). Indeed, 5\% of GCs should be bright enough at formation to
  detect in the HUDF ($\muv = -17$) even if $\xi=1$. If $\xi=10$, 50\% of GCs
  reach $\muv=-17$ and 5\% reach $\muv = -19.5$ (albeit for only $\sim 5$ Myr).

  Eq.~\eqref{eq:observability} gives the number of GCs observable in a field
  with fixed angular size; I estimate that 20 (116) GCs are already visible,
  with $\muv < -17$, in the \textit{Hubble} Ultra Deep Field if $\xi=1$
  (10). \jwst\ will likely be sensitive to $\sim 150-1000$ GCs in a survey that
  reaches $\muv=-15$ ($m=31.7$ at $z=6$), depending on $\xi$. \textit{Given that
    only $\sim 150$ sources with $\muv(z\ga 6) < -17$ have been detected in the
    UDF, I find that values of $\xi \ga 10$ are strongly disfavored.} The HUDF
  therefore already provides an important constraint on the formation of GCs and
  the origin of their abundance patterns; \jwst\ will almost certainly provide
  definitive data regarding $\xi$.
\item \textbf{Reionization}: even with $\xi \approx 5$, metal-poor globular
  clusters may drive cosmic reionization so long as they have relatively high
  escape fraction ($f_{\rm esc} \ga 0.3$), as has previously been argued by
  \citet{ricotti2002, schaerer2011, katz2013}.
\item \textbf{Constraints on dark matter models}: Many of the faintest sources
  at high redshifts in \hst\ lensing fields appear to be extremely compact
  \citep{vanzella2017, bouwens2017}. If some non-negligible fraction of these
  sources are GCs in formation, then the constraints on dark matter models
  obtained by comparing the observed number density of galaxies to the predicted
  number density of dark matter halos are weakened substantially: observed
  sources may be bright proto-GCs in more massive halos rather than ``typical''
  galaxies in less massive halos.
\item \textbf{Halo masses at low redshifts}: The faintest galaxies hosting at
  least one globular cluster (e.g., Pegasus and Eridanus II) are very likely to
  be in halos with masses of
  $7\times 10^{9}\la \mhalo \la 3 \times 10^{10}\,\msun$ at the present day.
\item \textbf{Accreted versus in-situ clusters}: the fraction of metal-poor GCs
  formed in situ by $z=6$ varies strongly with halo mass, with all GC formation
  occurring in situ for the lowest-mass halos and most GCs being accreted for
  very massive (galaxy cluster) halos. Milky Way-mass systems form 17\% of their
  old GCs in situ (in the main progenitor by the end of GC formation), with a
  90\% confidence interval of 8-40\%. The typical accreted GC comes from a
  relatively low-mass progenitor hosting $\sim 2$ GCs. It is highly unlikely
  that all of the metal-poor GCs in the
  Fornax dSph were accreted in the past 10 Gyr.
\end{itemize}

Perhaps the most intriguing result of this paper is that \hst\ observations are
already constraining $\xi$ and that \jwst\ will likely provide a definitive test
of scenarios that require large populations of ``first-generation'' GC stars
($\xi \ga 10$). Even relatively low values of $3 \la \xi \la 5$ imply that GCs
may be dominant contributors to reionization. The need for obtaining more
accurate ages of old globular clusters -- a time-honored astronomical endeavor
-- is therefore pressing. As an example, accurate absolute age determinations
with a precision of $\la 0.8$ Gyr would be able to differentiate consider two
models, one in which globular clusters form at $6 \ga z \ga 3$
($0.93 \la t_{\rm cosmic} \la 2.14$ Gyr) and another with a formation epoch of
$10 \la z \la 6$ ($0.47 \la t_{\rm cosmic} \la 0.93$ Gyr). 

I have not considered the effects of globular cluster disruption (e.g.,
\citealt{spitzer1987, vesperini1997, gnedin1999b, trenti2010a}) in this
work. The motivation for this -- beyond convenience -- is that it provides a
conservative estimate on the observability of GCs in the reionization era and
their effects on dark matter constraints. A future paper will include the
effects of GC disruption and resulting implications for a variety of quantities,
including the contribution of disrupted GCs to the stellar halos of galaxies
(e.g., \citealt{boley2009}). However, immediate constraints on disruption come
from the scatter in the $\Mgcs-\mhalo$ relation. In particular, the mass
dependence of the \citet{harris2017} calibration of $\etab$ is negligible,
implying that either disruption is a minimal effect or is independent of halo
mass (which would be unexpected, given the strong dependence of galaxy mass on
halo mass). If disruption is indeed uniform across halo mass, the primary effect
would be to modify the derived efficiency of conversion of baryons into globular
clusters without affecting $\mmin$; the results presented in
Section~\ref{subsec:efficiency} are a lower limit. Such disruption would also
affect the $\mhalo-\Mgcs$ relation at $z=6$, but not the relation at $z=0$ and
could be fully incorporated into the present model through a parameter
$f_{\rm disrupt}$ via $\xi \rightarrow f_{\rm disrupt}\,\xi$. Dissolution owing
to two-body relaxation is likely to be an important effect for
$\mgc \ll 2\times 10^5\,\msun$ (i.e., below the characteristic mass of the $z=0$
globular cluster luminosity function; \citealt{fall2001}). However, $\Mgcs$ is
dominated by clusters above the characteristic mass, implying that evaporation
of low-mass globular clusters is at most a second-order effect in establishing
the observed $\Mgcs-\mhalo$ relation.

The model presented here is explicitly aimed at understanding metal-poor (blue)
globular clusters. This is motivated by the overall dominance of metal-poor
globular clusters over metal-rich ones and their earlier formation epoch
\citep{forbes2015}, which links them to the reionization era. Nevertheless, a
full model for globular cluster formation should also explain metal-rich
clusters and why they also scale linearly with halo mass
\citep{harris2015}. Merger-based formation models (e.g., \citealt{ashman1992,
  bekki2008}) are compelling in this context and will be the subject of forthcoming work.

A somewhat more speculative, but intriguing, future direction is to consider the
predictions of the current model in the context of supermassive black hole
formation. There is a well-established observational connection between the mass
of a central black hole and $\Mgcs$ for galaxies \citep{spitler2009,
  burkert2010}. Given that there is a threshold dark matter halo mass for GC
formation, is there an equivalent for supermassive black hole formation, and if
so, is the threshold mass the same (as would be implied by the observed
$M_{\rm BH}-\Mgcs$ relation)?  What would the implications be for black holes in
dwarf galaxies (e.g., \citealt{silk2017})? A detailed understanding of globular
cluster formation, fully embedded within a cosmological context, will have
far-reaching implications for diverse areas of astrophysics and cosmology.

\section*{Acknowledgments} 
I am very grateful to Yu Lu for providing his version of the Parkinson \& Cole
merger tree algorithm. I have benefitted from conversations with Rychard
Bouwens, James Bullock, Steve Finkelstein, Bill Harris, Mike Hudson, Pawan
Kumar, Rachael Livermore, Milos Milosavljevic, Eliot Quataert, Chris Sneden,
David Spergel, and Dan Weisz. The Near/Far Workshop in Santa Rosa in December
2016 provided motivation for me to develop some of the ideas presented
here. This work used python-fsps \citep{foreman-mackey2014}, and I thank Ben
Johnson for his help in using python-fsps.

Support for this work was provided by the National Science Foundation (grant
AST-1517226) and by NASA through grant NNX17AG29G and HST grants AR-12836,
AR-13888, AR-13896, GO-14191, and AR-14282 awarded by the Space Telescope
Science Institute, which is operated by the Association of Universities for
Research in Astronomy, Inc., under NASA contract NAS5-26555. Much of the
analysis in this paper relied on the python packages {\tt NumPy} \citep{numpy},
{\tt SciPy} \citep{scipy}, {\tt Matplotlib} \citep{matplotlib}, and {\tt
  iPython} \citep{ipython}, as well as {\tt pyfof}
(\href{https://pypi.python.org/pypi/pyfof/}
{https://pypi.python.org/pypi/pyfof/}); I am very grateful to the developers
of these tools. This research has made extensive use of NASA's Astrophysics Data
System (\href{http://adsabs.harvard.edu/}{http://adsabs.harvard.edu/}) and the
arXiv eprint service (\href{http://arxiv.org}{http://arxiv.org}).

\bibliography{/Users/mbk/Work/Misc/main_bib.bib}

\begin{thebibliography}{}
\makeatletter
\relax
\def\mn@urlcharsother{\let\do\@makeother \do\$\do\&\do\#\do\^\do\_\do\%\do\~}
\def\mn@doi{\begingroup\mn@urlcharsother \@ifnextchar [ {\mn@doi@}
  {\mn@doi@[]}}
\def\mn@doi@[#1]#2{\def\@tempa{#1}\ifx\@tempa\@empty \href
  {http://dx.doi.org/#2} {doi:#2}\else \href {http://dx.doi.org/#2} {#1}\fi
  \endgroup}
\def\mn@eprint#1#2{\mn@eprint@#1:#2::\@nil}
\def\mn@eprint@arXiv#1{\href {http://arxiv.org/abs/#1} {{\tt arXiv:#1}}}
\def\mn@eprint@dblp#1{\href {http://dblp.uni-trier.de/rec/bibtex/#1.xml}
  {dblp:#1}}
\def\mn@eprint@#1:#2:#3:#4\@nil{\def\@tempa {#1}\def\@tempb {#2}\def\@tempc
  {#3}\ifx \@tempc \@empty \let \@tempc \@tempb \let \@tempb \@tempa \fi \ifx
  \@tempb \@empty \def\@tempb {arXiv}\fi \@ifundefined
  {mn@eprint@\@tempb}{\@tempb:\@tempc}{\expandafter \expandafter \csname
  mn@eprint@\@tempb\endcsname \expandafter{\@tempc}}}

\bibitem[\protect\citeauthoryear{{Ashman} \& {Zepf}}{{Ashman} \&
  {Zepf}}{1992}]{ashman1992}
{Ashman} K.~M.,  {Zepf} S.~E.,  1992, \mn@doi [\apj] {10.1086/170850}, \href
  {http://adsabs.harvard.edu/abs/1992ApJ...384...50A} {384, 50}

\bibitem[\protect\citeauthoryear{{Bastian} \& {Lardo}}{{Bastian} \&
  {Lardo}}{2015}]{bastian2015}
{Bastian} N.,  {Lardo} C.,  2015, \mn@doi [\mnras] {10.1093/mnras/stv1661},
  \href {http://adsabs.harvard.edu/abs/2015MNRAS.453..357B} {453, 357}

\bibitem[\protect\citeauthoryear{{Bekki}, {Campbell}, {Lattanzio}  \&
  {Norris}}{{Bekki} et~al.}{2007}]{bekki2007}
{Bekki} K.,  {Campbell} S.~W.,  {Lattanzio} J.~C.,   {Norris} J.~E.,  2007,
  \mn@doi [\mnras] {10.1111/j.1365-2966.2007.11606.x}, \href
  {http://adsabs.harvard.edu/abs/2007MNRAS.377..335B} {377, 335}

\bibitem[\protect\citeauthoryear{{Bekki}, {Yahagi}, {Nagashima}  \&
  {Forbes}}{{Bekki} et~al.}{2008}]{bekki2008}
{Bekki} K.,  {Yahagi} H.,  {Nagashima} M.,   {Forbes} D.~A.,  2008, \mn@doi
  [\mnras] {10.1111/j.1365-2966.2008.13318.x}, \href
  {http://adsabs.harvard.edu/abs/2008MNRAS.387.1131B} {387, 1131}

\bibitem[\protect\citeauthoryear{{Blakeslee}, {Tonry}  \&
  {Metzger}}{{Blakeslee} et~al.}{1997}]{blakeslee1997}
{Blakeslee} J.~P.,  {Tonry} J.~L.,   {Metzger} M.~R.,  1997, \mn@doi [\aj]
  {10.1086/118488}, \href {http://adsabs.harvard.edu/abs/1997AJ....114..482B}
  {114, 482}

\bibitem[\protect\citeauthoryear{{Boley}, {Lake}, {Read}  \&
  {Teyssier}}{{Boley} et~al.}{2009}]{boley2009}
{Boley} A.~C.,  {Lake} G.,  {Read} J.,   {Teyssier} R.,  2009, \mn@doi [\apjl]
  {10.1088/0004-637X/706/1/L192}, \href
  {http://adsabs.harvard.edu/abs/2009ApJ...706L.192B} {706, L192}

\bibitem[\protect\citeauthoryear{{Bouwens} et~al.,}{{Bouwens}
  et~al.}{2015}]{bouwens2015}
{Bouwens} R.~J.,  et~al., 2015, \mn@doi [\apj] {10.1088/0004-637X/803/1/34},
  \href {http://adsabs.harvard.edu/abs/2015ApJ...803...34B} {803, 34}

\bibitem[\protect\citeauthoryear{{Bouwens}, {Illingworth}, {Oesch}, {Atek},
  {Lam}  \& {Stefanon}}{{Bouwens} et~al.}{2017}]{bouwens2017}
{Bouwens} R.~J.,  {Illingworth} G.~D.,  {Oesch} P.~A.,  {Atek} H.,  {Lam} D.,
  {Stefanon} M.,  2017, \mn@doi [\apj] {10.3847/1538-4357/aa74e4}, \href
  {http://adsabs.harvard.edu/abs/2017ApJ...843...41B} {843, 41}

\bibitem[\protect\citeauthoryear{{Boylan-Kolchin}, {Springel}, {White}  \&
  {Jenkins}}{{Boylan-Kolchin} et~al.}{2010}]{boylan-kolchin2010}
{Boylan-Kolchin} M.,  {Springel} V.,  {White} S.~D.~M.,   {Jenkins} A.,  2010,
  \mn@doi [\mnras] {10.1111/j.1365-2966.2010.16774.x}, \href
  {http://adsabs.harvard.edu/abs/2010MNRAS.406..896B} {406, 896}

\bibitem[\protect\citeauthoryear{{Boylan-Kolchin}, {Bullock}  \&
  {Kaplinghat}}{{Boylan-Kolchin} et~al.}{2012}]{boylan-kolchin2012}
{Boylan-Kolchin} M.,  {Bullock} J.~S.,   {Kaplinghat} M.,  2012, \mn@doi
  [\mnras] {10.1111/j.1365-2966.2012.20695.x}, \href
  {http://adsabs.harvard.edu/abs/2012MNRAS.422.1203B} {422, 1203}

\bibitem[\protect\citeauthoryear{{Boylan-Kolchin}, {Bullock}  \&
  {Garrison-Kimmel}}{{Boylan-Kolchin} et~al.}{2014}]{boylan-kolchin2014}
{Boylan-Kolchin} M.,  {Bullock} J.~S.,   {Garrison-Kimmel} S.,  2014, \mn@doi
  [\mnras] {10.1093/mnrasl/slu079}, \href
  {http://adsabs.harvard.edu/abs/2014MNRAS.443L..44B} {443, L44}

\bibitem[\protect\citeauthoryear{{Boylan-Kolchin}, {Weisz}, {Johnson},
  {Bullock}, {Conroy}  \& {Fitts}}{{Boylan-Kolchin}
  et~al.}{2015}]{boylan-kolchin2015}
{Boylan-Kolchin} M.,  {Weisz} D.~R.,  {Johnson} B.~D.,  {Bullock} J.~S.,
  {Conroy} C.,   {Fitts} A.,  2015, \mn@doi [\mnras] {10.1093/mnras/stv1736},
  \href {http://adsabs.harvard.edu/abs/2015MNRAS.453.1503B} {453, 1503}

\bibitem[\protect\citeauthoryear{{Boylan-Kolchin}, {Weisz}, {Bullock}  \&
  {Cooper}}{{Boylan-Kolchin} et~al.}{2016}]{boylan-kolchin2016}
{Boylan-Kolchin} M.,  {Weisz} D.~R.,  {Bullock} J.~S.,   {Cooper} M.~C.,  2016,
  \mn@doi [\mnras] {10.1093/mnrasl/slw121}, \href
  {http://adsabs.harvard.edu/abs/2016MNRAS.462L..51B} {462, L51}

\bibitem[\protect\citeauthoryear{{Bozek}, {Marsh}, {Silk}  \& {Wyse}}{{Bozek}
  et~al.}{2015}]{bozek2015}
{Bozek} B.,  {Marsh} D.~J.~E.,  {Silk} J.,   {Wyse} R.~F.~G.,  2015, \mn@doi
  [\mnras] {10.1093/mnras/stv624}, \href
  {http://adsabs.harvard.edu/abs/2015MNRAS.450..209B} {450, 209}

\bibitem[\protect\citeauthoryear{{Brodie} \& {Strader}}{{Brodie} \&
  {Strader}}{2006}]{brodie2006}
{Brodie} J.~P.,  {Strader} J.,  2006, \mn@doi [\araa]
  {10.1146/annurev.astro.44.051905.092441}, \href
  {http://adsabs.harvard.edu/abs/2006ARA%26A..44..193B} {44, 193}

\bibitem[\protect\citeauthoryear{{Bromm} \& {Clarke}}{{Bromm} \&
  {Clarke}}{2002}]{bromm2002}
{Bromm} V.,  {Clarke} C.~J.,  2002, \mn@doi [\apjl] {10.1086/339440}, \href
  {http://adsabs.harvard.edu/abs/2002ApJ...566L...1B} {566, L1}

\bibitem[\protect\citeauthoryear{{Burkert} \& {Tremaine}}{{Burkert} \&
  {Tremaine}}{2010}]{burkert2010}
{Burkert} A.,  {Tremaine} S.,  2010, \mn@doi [\apj]
  {10.1088/0004-637X/720/1/516}, \href
  {http://adsabs.harvard.edu/abs/2010ApJ...720..516B} {720, 516}

\bibitem[\protect\citeauthoryear{{Byler}, {Dalcanton}, {Conroy}  \&
  {Johnson}}{{Byler} et~al.}{2017}]{byler2017}
{Byler} N.,  {Dalcanton} J.~J.,  {Conroy} C.,   {Johnson} B.~D.,  2017, \mn@doi
  [\apj] {10.3847/1538-4357/aa6c66}, \href
  {http://adsabs.harvard.edu/abs/2017ApJ...840...44B} {840, 44}

\bibitem[\protect\citeauthoryear{{Cen}}{{Cen}}{2001}]{cen2001}
{Cen} R.,  2001, \mn@doi [\apj] {10.1086/323071}, \href
  {http://adsabs.harvard.edu/abs/2001ApJ...560..592C} {560, 592}

\bibitem[\protect\citeauthoryear{{Choi}, {Dotter}, {Conroy}, {Cantiello},
  {Paxton}  \& {Johnson}}{{Choi} et~al.}{2016}]{choi2016}
{Choi} J.,  {Dotter} A.,  {Conroy} C.,  {Cantiello} M.,  {Paxton} B.,
  {Johnson} B.~D.,  2016, \mn@doi [\apj] {10.3847/0004-637X/823/2/102}, \href
  {http://adsabs.harvard.edu/abs/2016ApJ...823..102C} {823, 102}

\bibitem[\protect\citeauthoryear{{Cole}, {Dehnen}, {Read}  \&
  {Wilkinson}}{{Cole} et~al.}{2012}]{cole2012}
{Cole} D.~R.,  {Dehnen} W.,  {Read} J.~I.,   {Wilkinson} M.~I.,  2012, \mn@doi
  [\mnras] {10.1111/j.1365-2966.2012.21885.x}, \href
  {http://adsabs.harvard.edu/abs/2012MNRAS.426..601C} {426, 601}

\bibitem[\protect\citeauthoryear{{Cole} et~al.,}{{Cole}
  et~al.}{2017}]{cole2017}
{Cole} A.~A.,  et~al., 2017, \mn@doi [\apj] {10.3847/1538-4357/aa5df6}, \href
  {http://adsabs.harvard.edu/abs/2017ApJ...837...54C} {837, 54}

\bibitem[\protect\citeauthoryear{{Conroy}}{{Conroy}}{2012}]{conroy2012a}
{Conroy} C.,  2012, \mn@doi [\apj] {10.1088/0004-637X/758/1/21}, \href
  {http://adsabs.harvard.edu/abs/2012ApJ...758...21C} {758, 21}

\bibitem[\protect\citeauthoryear{{Conroy} \& {Gunn}}{{Conroy} \&
  {Gunn}}{2010}]{conroy2010a}
{Conroy} C.,  {Gunn} J.~E.,  2010, \mn@doi [\apj]
  {10.1088/0004-637X/712/2/833}, \href
  {http://adsabs.harvard.edu/abs/2010ApJ...712..833C} {712, 833}

\bibitem[\protect\citeauthoryear{{Conroy}, {Gunn}  \& {White}}{{Conroy}
  et~al.}{2009}]{conroy2009a}
{Conroy} C.,  {Gunn} J.~E.,   {White} M.,  2009, \mn@doi [\apj]
  {10.1088/0004-637X/699/1/486}, \href
  {http://adsabs.harvard.edu/abs/2009ApJ...699..486C} {699, 486}

\bibitem[\protect\citeauthoryear{{Conroy}, {Loeb}  \& {Spergel}}{{Conroy}
  et~al.}{2011}]{conroy2011a}
{Conroy} C.,  {Loeb} A.,   {Spergel} D.~N.,  2011, \mn@doi [\apj]
  {10.1088/0004-637X/741/2/72}, \href
  {http://adsabs.harvard.edu/abs/2011ApJ...741...72C} {741, 72}

\bibitem[\protect\citeauthoryear{{Corbett Moran}, {Teyssier}  \&
  {Lake}}{{Corbett Moran} et~al.}{2014}]{corbett-moran2014}
{Corbett Moran} C.,  {Teyssier} R.,   {Lake} G.,  2014, \mn@doi [\mnras]
  {10.1093/mnras/stu1057}, \href
  {http://adsabs.harvard.edu/abs/2014MNRAS.442.2826C} {442, 2826}

\bibitem[\protect\citeauthoryear{{Crnojevi{\'c}}, {Sand}, {Zaritsky},
  {Spekkens}, {Willman}  \& {Hargis}}{{Crnojevi{\'c}}
  et~al.}{2016}]{crnojevic2016}
{Crnojevi{\'c}} D.,  {Sand} D.~J.,  {Zaritsky} D.,  {Spekkens} K.,  {Willman}
  B.,   {Hargis} J.~R.,  2016, \mn@doi [\apjl] {10.3847/2041-8205/824/1/L14},
  \href {http://adsabs.harvard.edu/abs/2016ApJ...824L..14C} {824, L14}

\bibitem[\protect\citeauthoryear{{D'Ercole}, {Vesperini}, {D'Antona},
  {McMillan}  \& {Recchi}}{{D'Ercole} et~al.}{2008}]{dercole2008}
{D'Ercole} A.,  {Vesperini} E.,  {D'Antona} F.,  {McMillan} S.~L.~W.,
  {Recchi} S.,  2008, \mn@doi [\mnras] {10.1111/j.1365-2966.2008.13915.x},
  \href {http://adsabs.harvard.edu/abs/2008MNRAS.391..825D} {391, 825}

\bibitem[\protect\citeauthoryear{{Dotter}}{{Dotter}}{2016}]{dotter2016}
{Dotter} A.,  2016, \mn@doi [\apjs] {10.3847/0067-0049/222/1/8}, \href
  {http://adsabs.harvard.edu/abs/2016ApJS..222....8D} {222, 8}

\bibitem[\protect\citeauthoryear{{Elmegreen}, {Malhotra}  \&
  {Rhoads}}{{Elmegreen} et~al.}{2012}]{elmegreen2012}
{Elmegreen} B.~G.,  {Malhotra} S.,   {Rhoads} J.,  2012, \mn@doi [\apj]
  {10.1088/0004-637X/757/1/9}, \href
  {http://adsabs.harvard.edu/abs/2012ApJ...757....9E} {757, 9}

\bibitem[\protect\citeauthoryear{{Fall} \& {Rees}}{{Fall} \&
  {Rees}}{1985}]{fall1985}
{Fall} S.~M.,  {Rees} M.~J.,  1985, \mn@doi [\apj] {10.1086/163585}, \href
  {http://adsabs.harvard.edu/abs/1985ApJ...298...18F} {298, 18}

\bibitem[\protect\citeauthoryear{{Fall} \& {Zhang}}{{Fall} \&
  {Zhang}}{2001}]{fall2001}
{Fall} S.~M.,  {Zhang} Q.,  2001, \mn@doi [\apj] {10.1086/323358}, \href
  {http://adsabs.harvard.edu/abs/2001ApJ...561..751F} {561, 751}

\bibitem[\protect\citeauthoryear{{Finkelstein} et~al.,}{{Finkelstein}
  et~al.}{2015}]{finkelstein2015}
{Finkelstein} S.~L.,  et~al., 2015, \mn@doi [\apj]
  {10.1088/0004-637X/810/1/71}, \href
  {http://adsabs.harvard.edu/abs/2015ApJ...810...71F} {810, 71}

\bibitem[\protect\citeauthoryear{{Forbes}, {Pastorello}, {Romanowsky}, {Usher},
  {Brodie}  \& {Strader}}{{Forbes} et~al.}{2015}]{forbes2015}
{Forbes} D.~A.,  {Pastorello} N.,  {Romanowsky} A.~J.,  {Usher} C.,  {Brodie}
  J.~P.,   {Strader} J.,  2015, \mn@doi [\mnras] {10.1093/mnras/stv1312}, \href
  {http://adsabs.harvard.edu/abs/2015MNRAS.452.1045F} {452, 1045}

\bibitem[\protect\citeauthoryear{Foreman-Mackey, Sick  \&
  Johnson}{Foreman-Mackey et~al.}{2014}]{foreman-mackey2014}
Foreman-Mackey D.,  Sick J.,   Johnson B.,  2014, python-fsps: Python bindings
  to FSPS (v0.1.1), \url {https://doi.org/10.5281/zenodo.12157}

\bibitem[\protect\citeauthoryear{{Garrison-Kimmel}, {Bullock}, {Boylan-Kolchin}
   \& {Bardwell}}{{Garrison-Kimmel} et~al.}{2017}]{garrison-kimmel2017}
{Garrison-Kimmel} S.,  {Bullock} J.~S.,  {Boylan-Kolchin} M.,   {Bardwell} E.,
  2017, \mn@doi [\mnras] {10.1093/mnras/stw2564}, \href
  {http://adsabs.harvard.edu/abs/2017MNRAS.464.3108G} {464, 3108}

\bibitem[\protect\citeauthoryear{{Georgiev}, {Puzia}, {Hilker}  \&
  {Goudfrooij}}{{Georgiev} et~al.}{2009}]{georgiev2009}
{Georgiev} I.~Y.,  {Puzia} T.~H.,  {Hilker} M.,   {Goudfrooij} P.,  2009,
  \mn@doi [\mnras] {10.1111/j.1365-2966.2008.14104.x}, \href
  {http://adsabs.harvard.edu/abs/2009MNRAS.392..879G} {392, 879}

\bibitem[\protect\citeauthoryear{{Georgiev}, {Puzia}, {Goudfrooij}  \&
  {Hilker}}{{Georgiev} et~al.}{2010}]{georgiev2010}
{Georgiev} I.~Y.,  {Puzia} T.~H.,  {Goudfrooij} P.,   {Hilker} M.,  2010,
  \mn@doi [\mnras] {10.1111/j.1365-2966.2010.16802.x}, \href
  {http://adsabs.harvard.edu/abs/2010MNRAS.406.1967G} {406, 1967}

\bibitem[\protect\citeauthoryear{{Girardi} et~al.,}{{Girardi}
  et~al.}{2010}]{girardi2010}
{Girardi} L.,  et~al., 2010, \mn@doi [\apj] {10.1088/0004-637X/724/2/1030},
  \href {http://adsabs.harvard.edu/abs/2010ApJ...724.1030G} {724, 1030}

\bibitem[\protect\citeauthoryear{{Gnedin}, {Lee}  \& {Ostriker}}{{Gnedin}
  et~al.}{1999}]{gnedin1999b}
{Gnedin} O.~Y.,  {Lee} H.~M.,   {Ostriker} J.~P.,  1999, \mn@doi [\apj]
  {10.1086/307659}, \href {http://adsabs.harvard.edu/abs/1999ApJ...522..935G}
  {522, 935}

\bibitem[\protect\citeauthoryear{{Goerdt}, {Moore}, {Read}, {Stadel}  \&
  {Zemp}}{{Goerdt} et~al.}{2006}]{goerdt2006}
{Goerdt} T.,  {Moore} B.,  {Read} J.~I.,  {Stadel} J.,   {Zemp} M.,  2006,
  \mn@doi [\mnras] {10.1111/j.1365-2966.2006.10182.x}, \href
  {http://adsabs.harvard.edu/abs/2006MNRAS.368.1073G} {368, 1073}

\bibitem[\protect\citeauthoryear{{Gratton}, {Carretta}  \&
  {Bragaglia}}{{Gratton} et~al.}{2012}]{gratton2012}
{Gratton} R.~G.,  {Carretta} E.,   {Bragaglia} A.,  2012, \mn@doi [\aapr]
  {10.1007/s00159-012-0050-3}, \href
  {http://adsabs.harvard.edu/abs/2012A%26ARv..20...50G} {20, 50}

\bibitem[\protect\citeauthoryear{{Gunn}}{{Gunn}}{1980}]{gunn1980}
{Gunn} J.~E.,  1980, in {Hanes} D.,  {Madore} B.,  eds, Globular Clusters.
  p.~301

\bibitem[\protect\citeauthoryear{{Harris}, {Harris}  \& {Alessi}}{{Harris}
  et~al.}{2013}]{harris2013}
{Harris} W.~E.,  {Harris} G.~L.~H.,   {Alessi} M.,  2013, \mn@doi [\apj]
  {10.1088/0004-637X/772/2/82}, \href
  {http://adsabs.harvard.edu/abs/2013ApJ...772...82H} {772, 82}

\bibitem[\protect\citeauthoryear{{Harris}, {Harris}  \& {Hudson}}{{Harris}
  et~al.}{2015}]{harris2015}
{Harris} W.~E.,  {Harris} G.~L.,   {Hudson} M.~J.,  2015, \mn@doi [\apj]
  {10.1088/0004-637X/806/1/36}, \href
  {http://adsabs.harvard.edu/abs/2015ApJ...806...36H} {806, 36}

\bibitem[\protect\citeauthoryear{{Harris}, {Blakeslee}  \& {Harris}}{{Harris}
  et~al.}{2017}]{harris2017}
{Harris} W.~E.,  {Blakeslee} J.~P.,   {Harris} G.~L.~H.,  2017, \mn@doi [\apj]
  {10.3847/1538-4357/836/1/67}, \href
  {http://adsabs.harvard.edu/abs/2017ApJ...836...67H} {836, 67}

\bibitem[\protect\citeauthoryear{{Hu}, {Barkana}  \& {Gruzinov}}{{Hu}
  et~al.}{2000}]{hu2000}
{Hu} W.,  {Barkana} R.,   {Gruzinov} A.,  2000, \mn@doi [\prl]
  {10.1103/PhysRevLett.85.1158}, \href
  {http://adsabs.harvard.edu/abs/2000PhRvL..85.1158H} {85, 1158}

\bibitem[\protect\citeauthoryear{{Hudson}, {Harris}  \& {Harris}}{{Hudson}
  et~al.}{2014}]{hudson2014}
{Hudson} M.~J.,  {Harris} G.~L.,   {Harris} W.~E.,  2014, \mn@doi [\apjl]
  {10.1088/2041-8205/787/1/L5}, \href
  {http://adsabs.harvard.edu/abs/2014ApJ...787L...5H} {787, L5}

\bibitem[\protect\citeauthoryear{{Hui}, {Ostriker}, {Tremaine}  \&
  {Witten}}{{Hui} et~al.}{2017}]{hui2017}
{Hui} L.,  {Ostriker} J.~P.,  {Tremaine} S.,   {Witten} E.,  2017, \mn@doi
  [\prd] {10.1103/PhysRevD.95.043541}, \href
  {http://adsabs.harvard.edu/abs/2017PhRvD..95d3541H} {95, 043541}

\bibitem[\protect\citeauthoryear{Hunter}{Hunter}{2007}]{matplotlib}
Hunter J.~D.,  2007, Computing In Science \& Engineering, 9, 90

\bibitem[\protect\citeauthoryear{Jones, Oliphant, Peterson  et~al.}{Jones
  et~al.}{2001}]{scipy}
Jones E.,  Oliphant T.,  Peterson P.,   et~al., 2001, {SciPy}: Open source
  scientific tools for {Python}, \url {http://www.scipy.org/}

\bibitem[\protect\citeauthoryear{{Kang}, {Shapiro}, {Fall}  \& {Rees}}{{Kang}
  et~al.}{1990}]{kang1990}
{Kang} H.,  {Shapiro} P.~R.,  {Fall} S.~M.,   {Rees} M.~J.,  1990, \mn@doi
  [\apj] {10.1086/169360}, \href
  {http://adsabs.harvard.edu/abs/1990ApJ...363..488K} {363, 488}

\bibitem[\protect\citeauthoryear{{Katz} \& {Ricotti}}{{Katz} \&
  {Ricotti}}{2013}]{katz2013}
{Katz} H.,  {Ricotti} M.,  2013, \mn@doi [\mnras] {10.1093/mnras/stt676}, \href
  {http://adsabs.harvard.edu/abs/2013MNRAS.432.3250K} {432, 3250}

\bibitem[\protect\citeauthoryear{{Katz} \& {Ricotti}}{{Katz} \&
  {Ricotti}}{2014}]{katz2014}
{Katz} H.,  {Ricotti} M.,  2014, \mn@doi [\mnras] {10.1093/mnras/stu1489},
  \href {http://adsabs.harvard.edu/abs/2014MNRAS.444.2377K} {444, 2377}

\bibitem[\protect\citeauthoryear{{Kennicutt}}{{Kennicutt}}{1998}]{kennicutt1998}
{Kennicutt} Jr. R.~C.,  1998, \mn@doi [\araa] {10.1146/annurev.astro.36.1.189},
  \href {http://adsabs.harvard.edu/abs/1998ARA%26A..36..189K} {36, 189}

\bibitem[\protect\citeauthoryear{{Kimm}, {Cen}, {Rosdahl}  \& {Yi}}{{Kimm}
  et~al.}{2016}]{kimm2016}
{Kimm} T.,  {Cen} R.,  {Rosdahl} J.,   {Yi} S.~K.,  2016, \mn@doi [\apj]
  {10.3847/0004-637X/823/1/52}, \href
  {http://adsabs.harvard.edu/abs/2016ApJ...823...52K} {823, 52}

\bibitem[\protect\citeauthoryear{{Koposov}, {Belokurov}, {Torrealba}  \&
  {Evans}}{{Koposov} et~al.}{2015}]{koposov2015}
{Koposov} S.~E.,  {Belokurov} V.,  {Torrealba} G.,   {Evans} N.~W.,  2015,
  \mn@doi [\apj] {10.1088/0004-637X/805/2/130}, \href
  {http://adsabs.harvard.edu/abs/2015ApJ...805..130K} {805, 130}

\bibitem[\protect\citeauthoryear{{Kravtsov} \& {Gnedin}}{{Kravtsov} \&
  {Gnedin}}{2005}]{kravtsov2005}
{Kravtsov} A.~V.,  {Gnedin} O.~Y.,  2005, \mn@doi [\apj] {10.1086/428636},
  \href {http://adsabs.harvard.edu/abs/2005ApJ...623..650K} {623, 650}

\bibitem[\protect\citeauthoryear{{Kroupa}}{{Kroupa}}{2001}]{kroupa2001}
{Kroupa} P.,  2001, \mn@doi [\mnras] {10.1046/j.1365-8711.2001.04022.x}, \href
  {http://adsabs.harvard.edu/abs/2001MNRAS.322..231K} {322, 231}

\bibitem[\protect\citeauthoryear{{Kruijssen}}{{Kruijssen}}{2015}]{kruijssen2015}
{Kruijssen} J.~M.~D.,  2015, \mn@doi [\mnras] {10.1093/mnras/stv2026}, \href
  {http://adsabs.harvard.edu/abs/2015MNRAS.454.1658K} {454, 1658}

\bibitem[\protect\citeauthoryear{{Kuhlen}}{{Kuhlen}}{2010}]{kuhlen2010}
{Kuhlen} M.,  2010, \mn@doi [Advances in Astronomy] {10.1155/2010/162083},
  \href {http://adsabs.harvard.edu/abs/2010AdAst2010E..45K} {2010}

\bibitem[\protect\citeauthoryear{{Kuhlen} \& {Faucher-Gigu{\`e}re}}{{Kuhlen} \&
  {Faucher-Gigu{\`e}re}}{2012}]{kuhlen2012a}
{Kuhlen} M.,  {Faucher-Gigu{\`e}re} C.-A.,  2012, \mn@doi [\mnras]
  {10.1111/j.1365-2966.2012.20924.x}, \href
  {http://adsabs.harvard.edu/abs/2012MNRAS.423..862K} {423, 862}

\bibitem[\protect\citeauthoryear{{Lacey} \& {Cole}}{{Lacey} \&
  {Cole}}{1993}]{lacey1993}
{Lacey} C.,  {Cole} S.,  1993, \mnras, \href
  {http://adsabs.harvard.edu/abs/1993MNRAS.262..627L} {262, 627}

\bibitem[\protect\citeauthoryear{{Ma}, {Kasen}, {Hopkins},
  {Faucher-Gigu{\`e}re}, {Quataert}, {Kere{\v s}}  \& {Murray}}{{Ma}
  et~al.}{2015}]{ma2015}
{Ma} X.,  {Kasen} D.,  {Hopkins} P.~F.,  {Faucher-Gigu{\`e}re} C.-A.,
  {Quataert} E.,  {Kere{\v s}} D.,   {Murray} N.,  2015, \mn@doi [\mnras]
  {10.1093/mnras/stv1679}, \href
  {http://adsabs.harvard.edu/abs/2015MNRAS.453..960M} {453, 960}

\bibitem[\protect\citeauthoryear{{Mackey} \& {Gilmore}}{{Mackey} \&
  {Gilmore}}{2003}]{mackey2003}
{Mackey} A.~D.,  {Gilmore} G.~F.,  2003, \mn@doi [\mnras]
  {10.1046/j.1365-8711.2003.06275.x}, \href
  {http://adsabs.harvard.edu/abs/2003MNRAS.340..175M} {340, 175}

\bibitem[\protect\citeauthoryear{{Madau} \& {Dickinson}}{{Madau} \&
  {Dickinson}}{2014}]{madau2014}
{Madau} P.,  {Dickinson} M.,  2014, \mn@doi [\araa]
  {10.1146/annurev-astro-081811-125615}, \href
  {http://adsabs.harvard.edu/abs/2014ARA%26A..52..415M} {52, 415}

\bibitem[\protect\citeauthoryear{{Madau}, {Haardt}  \& {Rees}}{{Madau}
  et~al.}{1999}]{madau1999}
{Madau} P.,  {Haardt} F.,   {Rees} M.~J.,  1999, \mn@doi [\apj]
  {10.1086/306975}, \href {http://adsabs.harvard.edu/abs/1999ApJ...514..648M}
  {514, 648}

\bibitem[\protect\citeauthoryear{{Marigo}, {Girardi}, {Bressan}, {Groenewegen},
  {Silva}  \& {Granato}}{{Marigo} et~al.}{2008}]{marigo2008}
{Marigo} P.,  {Girardi} L.,  {Bressan} A.,  {Groenewegen} M.~A.~T.,  {Silva}
  L.,   {Granato} G.~L.,  2008, \mn@doi [\aap] {10.1051/0004-6361:20078467},
  \href {http://adsabs.harvard.edu/abs/2008A%26A...482..883M} {482, 883}

\bibitem[\protect\citeauthoryear{{McCrea}}{{McCrea}}{1982}]{mccrea1982}
{McCrea} W.~H.,  1982, in {Wolfendale} A.~W.,  ed.,  Astrophysics and Space
  Science Library Vol. 99, Progress in Cosmology. pp 239--257,
  \mn@doi{10.1007/978-94-009-7873-7_17}

\bibitem[\protect\citeauthoryear{{McQuinn}}{{McQuinn}}{2016}]{mcquinn2016}
{McQuinn} M.,  2016, \mn@doi [\araa] {10.1146/annurev-astro-082214-122355},
  \href {http://adsabs.harvard.edu/abs/2016ARA%26A..54..313M} {54, 313}

\bibitem[\protect\citeauthoryear{{Menci}, {Merle}, {Totzauer}, {Schneider},
  {Grazian}, {Castellano}  \& {Sanchez}}{{Menci} et~al.}{2017}]{menci2017}
{Menci} N.,  {Merle} A.,  {Totzauer} M.,  {Schneider} A.,  {Grazian} A.,
  {Castellano} M.,   {Sanchez} N.~G.,  2017, \mn@doi [\apj]
  {10.3847/1538-4357/836/1/61}, \href
  {http://adsabs.harvard.edu/abs/2017ApJ...836...61M} {836, 61}

\bibitem[\protect\citeauthoryear{{Mirabel}, {Dijkstra}, {Laurent}, {Loeb}  \&
  {Pritchard}}{{Mirabel} et~al.}{2011}]{mirabel2011}
{Mirabel} I.~F.,  {Dijkstra} M.,  {Laurent} P.,  {Loeb} A.,   {Pritchard}
  J.~R.,  2011, \mn@doi [\aap] {10.1051/0004-6361/201016357}, \href
  {http://adsabs.harvard.edu/abs/2011A%26A...528A.149M} {528, A149}

\bibitem[\protect\citeauthoryear{{Moore}}{{Moore}}{1996}]{moore1996a}
{Moore} B.,  1996, \mn@doi [\apjl] {10.1086/309998}, \href
  {http://adsabs.harvard.edu/abs/1996ApJ...461L..13M} {461, L13}

\bibitem[\protect\citeauthoryear{{Moore}, {Diemand}, {Madau}, {Zemp}  \&
  {Stadel}}{{Moore} et~al.}{2006}]{moore2006}
{Moore} B.,  {Diemand} J.,  {Madau} P.,  {Zemp} M.,   {Stadel} J.,  2006,
  \mn@doi [\mnras] {10.1111/j.1365-2966.2006.10116.x}, \href
  {http://adsabs.harvard.edu/abs/2006MNRAS.368..563M} {368, 563}

\bibitem[\protect\citeauthoryear{{Muratov} \& {Gnedin}}{{Muratov} \&
  {Gnedin}}{2010}]{muratov2010}
{Muratov} A.~L.,  {Gnedin} O.~Y.,  2010, \mn@doi [\apj]
  {10.1088/0004-637X/718/2/1266}, \href
  {http://adsabs.harvard.edu/abs/2010ApJ...718.1266M} {718, 1266}

\bibitem[\protect\citeauthoryear{{Murray} \& {Lin}}{{Murray} \&
  {Lin}}{1992}]{murray1992}
{Murray} S.~D.,  {Lin} D.~N.~C.,  1992, \mn@doi [\apj] {10.1086/171993}, \href
  {http://adsabs.harvard.edu/abs/1992ApJ...400..265M} {400, 265}

\bibitem[\protect\citeauthoryear{{Oh}, {Lin}  \& {Richer}}{{Oh}
  et~al.}{2000}]{oh2000}
{Oh} K.~S.,  {Lin} D.~N.~C.,   {Richer} H.~B.,  2000, \mn@doi [\apj]
  {10.1086/308477}, \href {http://adsabs.harvard.edu/abs/2000ApJ...531..727O}
  {531, 727}

\bibitem[\protect\citeauthoryear{{Parkinson}, {Cole}  \& {Helly}}{{Parkinson}
  et~al.}{2008}]{parkinson2008}
{Parkinson} H.,  {Cole} S.,   {Helly} J.,  2008, \mn@doi [\mnras]
  {10.1111/j.1365-2966.2007.12517.x}, \href
  {http://adsabs.harvard.edu/abs/2008MNRAS.383..557P} {383, 557}

\bibitem[\protect\citeauthoryear{{Pe{\~n}arrubia}, {McConnachie}  \&
  {Navarro}}{{Pe{\~n}arrubia} et~al.}{2008}]{penarrubia2008}
{Pe{\~n}arrubia} J.,  {McConnachie} A.~W.,   {Navarro} J.~F.,  2008, \mn@doi
  [\apj] {10.1086/521543}, \href
  {http://adsabs.harvard.edu/abs/2008ApJ...672..904P} {672, 904}

\bibitem[\protect\citeauthoryear{{Peebles}}{{Peebles}}{1984}]{peebles1984a}
{Peebles} P.~J.~E.,  1984, \mn@doi [\apj] {10.1086/161714}, \href
  {http://adsabs.harvard.edu/abs/1984ApJ...277..470P} {277, 470}

\bibitem[\protect\citeauthoryear{{Peebles} \& {Dicke}}{{Peebles} \&
  {Dicke}}{1968}]{peebles1968}
{Peebles} P.~J.~E.,  {Dicke} R.~H.,  1968, \mn@doi [\apj] {10.1086/149811},
  \href {http://adsabs.harvard.edu/abs/1968ApJ...154..891P} {154, 891}

\bibitem[\protect\citeauthoryear{P\'erez \& Granger}{P\'erez \&
  Granger}{2007}]{ipython}
P\'erez F.,  Granger B.~E.,  2007, \mn@doi [Computing in Science and
  Engineering] {10.1109/MCSE.2007.53}, 9, 21

\bibitem[\protect\citeauthoryear{{Planck Collaboration} et~al.,}{{Planck
  Collaboration} et~al.}{2016}]{planck2015}
{Planck Collaboration} et~al., 2016, \mn@doi [\aap]
  {10.1051/0004-6361/201525830}, \href
  {http://adsabs.harvard.edu/abs/2016A%26A...594A..13P} {594, A13}

\bibitem[\protect\citeauthoryear{{Popa}, {Naoz}, {Marinacci}  \&
  {Vogelsberger}}{{Popa} et~al.}{2016}]{popa2016}
{Popa} C.,  {Naoz} S.,  {Marinacci} F.,   {Vogelsberger} M.,  2016, \mn@doi
  [\mnras] {10.1093/mnras/stw1045}, \href
  {http://adsabs.harvard.edu/abs/2016MNRAS.460.1625P} {460, 1625}

\bibitem[\protect\citeauthoryear{{Press} \& {Schechter}}{{Press} \&
  {Schechter}}{1974}]{press1974}
{Press} W.~H.,  {Schechter} P.,  1974, \apj, \href
  {http://adsabs.harvard.edu/cgi-bin/nph-bib_query?bibcode=1974ApJ...187..425P&db_key=AST}
  {187, 425}

\bibitem[\protect\citeauthoryear{{Read}, {Iorio}, {Agertz}  \&
  {Fraternali}}{{Read} et~al.}{2017}]{read2017}
{Read} J.~I.,  {Iorio} G.,  {Agertz} O.,   {Fraternali} F.,  2017, \mn@doi
  [\mnras] {10.1093/mnras/stx147}, \href
  {http://adsabs.harvard.edu/abs/2017MNRAS.467.2019R} {467, 2019}

\bibitem[\protect\citeauthoryear{{Renzini}}{{Renzini}}{2017}]{renzini2017}
{Renzini} A.,  2017, \mn@doi [\mnras] {10.1093/mnrasl/slx057}, \href
  {http://adsabs.harvard.edu/abs/2017MNRAS.469L..63R} {469, L63}

\bibitem[\protect\citeauthoryear{{Renzini} et~al.,}{{Renzini}
  et~al.}{2015}]{renzini2015}
{Renzini} A.,  et~al., 2015, \mn@doi [\mnras] {10.1093/mnras/stv2268}, \href
  {http://adsabs.harvard.edu/abs/2015MNRAS.454.4197R} {454, 4197}

\bibitem[\protect\citeauthoryear{{Ricotti}}{{Ricotti}}{2002}]{ricotti2002}
{Ricotti} M.,  2002, \mn@doi [\mnras] {10.1046/j.1365-8711.2002.05990.x}, \href
  {http://adsabs.harvard.edu/abs/2002MNRAS.336L..33R} {336, L33}

\bibitem[\protect\citeauthoryear{{Robertson}, {Ellis}, {Furlanetto}  \&
  {Dunlop}}{{Robertson} et~al.}{2015}]{robertson2015}
{Robertson} B.~E.,  {Ellis} R.~S.,  {Furlanetto} S.~R.,   {Dunlop} J.~S.,
  2015, \mn@doi [\apjl] {10.1088/2041-8205/802/2/L19}, \href
  {http://adsabs.harvard.edu/abs/2015ApJ...802L..19R} {802, L19}

\bibitem[\protect\citeauthoryear{{Rosenblatt}, {Faber}  \&
  {Blumenthal}}{{Rosenblatt} et~al.}{1988}]{rosenblatt1988}
{Rosenblatt} E.~I.,  {Faber} S.~M.,   {Blumenthal} G.~R.,  1988, \mn@doi [\apj]
  {10.1086/166466}, \href {http://adsabs.harvard.edu/abs/1988ApJ...330..191R}
  {330, 191}

\bibitem[\protect\citeauthoryear{{Schaerer} \& {Charbonnel}}{{Schaerer} \&
  {Charbonnel}}{2011}]{schaerer2011}
{Schaerer} D.,  {Charbonnel} C.,  2011, \mn@doi [\mnras]
  {10.1111/j.1365-2966.2011.18304.x}, \href
  {http://adsabs.harvard.edu/abs/2011MNRAS.413.2297S} {413, 2297}

\bibitem[\protect\citeauthoryear{{Schultz}, {O{\~n}orbe}, {Abazajian}  \&
  {Bullock}}{{Schultz} et~al.}{2014}]{schultz2014}
{Schultz} C.,  {O{\~n}orbe} J.,  {Abazajian} K.~N.,   {Bullock} J.~S.,  2014,
  \mn@doi [\mnras] {10.1093/mnras/stu976}, \href
  {http://adsabs.harvard.edu/abs/2014MNRAS.442.1597S} {442, 1597}

\bibitem[\protect\citeauthoryear{{Sheth}, {Mo}  \& {Tormen}}{{Sheth}
  et~al.}{2001}]{sheth2001}
{Sheth} R.~K.,  {Mo} H.~J.,   {Tormen} G.,  2001, \mn@doi [\mnras]
  {10.1046/j.1365-8711.2001.04006.x}, \href
  {http://adsabs.harvard.edu/abs/2001MNRAS.323....1S} {323, 1}

\bibitem[\protect\citeauthoryear{Silk}{Silk}{2017}]{silk2017}
Silk J.,  2017, \apjl, 839, L13

\bibitem[\protect\citeauthoryear{{Spitler} \& {Forbes}}{{Spitler} \&
  {Forbes}}{2009}]{spitler2009}
{Spitler} L.~R.,  {Forbes} D.~A.,  2009, \mn@doi [\mnras]
  {10.1111/j.1745-3933.2008.00567.x}, \href
  {http://adsabs.harvard.edu/abs/2009MNRAS.392L...1S} {392, L1}

\bibitem[\protect\citeauthoryear{{Spitler}, {Romanowsky}, {Diemand}, {Strader},
  {Forbes}, {Moore}  \& {Brodie}}{{Spitler} et~al.}{2012}]{spitler2012}
{Spitler} L.~R.,  {Romanowsky} A.~J.,  {Diemand} J.,  {Strader} J.,  {Forbes}
  D.~A.,  {Moore} B.,   {Brodie} J.~P.,  2012, \mn@doi [\mnras]
  {10.1111/j.1365-2966.2012.21029.x}, \href
  {http://adsabs.harvard.edu/abs/2012MNRAS.423.2177S} {423, 2177}

\bibitem[\protect\citeauthoryear{{Spitzer}}{{Spitzer}}{1987}]{spitzer1987}
{Spitzer} L.,  1987, {Dynamical Evolution of Globular Clusters}.
Princeton, NJ, Princeton University Press

\bibitem[\protect\citeauthoryear{{Tremaine}}{{Tremaine}}{1976}]{tremaine1976a}
{Tremaine} S.~D.,  1976, \mn@doi [\apj] {10.1086/154085}, \href
  {http://adsabs.harvard.edu/abs/1976ApJ...203..345T} {203, 345}

\bibitem[\protect\citeauthoryear{{Trenti}, {Vesperini}  \& {Pasquato}}{{Trenti}
  et~al.}{2010}]{trenti2010a}
{Trenti} M.,  {Vesperini} E.,   {Pasquato} M.,  2010, \mn@doi [\apj]
  {10.1088/0004-637X/708/2/1598}, \href
  {http://adsabs.harvard.edu/abs/2010ApJ...708.1598T} {708, 1598}

\bibitem[\protect\citeauthoryear{{Van Der Walt}, {Colbert}  \&
  {Varoquaux}}{{Van Der Walt} et~al.}{2011}]{numpy}
{Van Der Walt} S.,  {Colbert} S.~C.,   {Varoquaux} G.,  2011, {arXiv:1102.1523
  [astro-ph]}, \href {http://adsabs.harvard.edu/abs/2011arXiv1102.1523V} {}

\bibitem[\protect\citeauthoryear{{VandenBerg}, {Brogaard}, {Leaman}  \&
  {Casagrande}}{{VandenBerg} et~al.}{2013}]{vandenberg2013}
{VandenBerg} D.~A.,  {Brogaard} K.,  {Leaman} R.,   {Casagrande} L.,  2013,
  \mn@doi [\apj] {10.1088/0004-637X/775/2/134}, \href
  {http://adsabs.harvard.edu/abs/2013ApJ...775..134V} {775, 134}

\bibitem[\protect\citeauthoryear{{Vanzella} et~al.,}{{Vanzella}
  et~al.}{2017}]{vanzella2017}
{Vanzella} E.,  et~al., 2017, \mn@doi [\mnras] {10.1093/mnras/stx351}, \href
  {http://adsabs.harvard.edu/abs/2017MNRAS.467.4304V} {467, 4304}

\bibitem[\protect\citeauthoryear{{Vesperini} \& {Heggie}}{{Vesperini} \&
  {Heggie}}{1997}]{vesperini1997}
{Vesperini} E.,  {Heggie} D.~C.,  1997, \mn@doi [\mnras]
  {10.1093/mnras/289.4.898}, \href
  {http://adsabs.harvard.edu/abs/1997MNRAS.289..898V} {289, 898}

\bibitem[\protect\citeauthoryear{{Weisz} \& {Boylan-Kolchin}}{{Weisz} \&
  {Boylan-Kolchin}}{2017}]{weisz2017}
{Weisz} D.~R.,  {Boylan-Kolchin} M.,  2017, \mn@doi [\mnras]
  {10.1093/mnrasl/slx043}, 469, L83

\bibitem[\protect\citeauthoryear{{Zaritsky}, {Crnojevi{\'c}}  \&
  {Sand}}{{Zaritsky} et~al.}{2016}]{zaritsky2016}
{Zaritsky} D.,  {Crnojevi{\'c}} D.,   {Sand} D.~J.,  2016, \mn@doi [\apjl]
  {10.3847/2041-8205/826/1/L9}, \href
  {http://adsabs.harvard.edu/abs/2016ApJ...826L...9Z} {826, L9}

\bibitem[\protect\citeauthoryear{{de Boer} \& {Fraser}}{{de Boer} \&
  {Fraser}}{2016}]{de-boer2016}
{de Boer} T.~J.~L.,  {Fraser} M.,  2016, \mn@doi [\aap]
  {10.1051/0004-6361/201527580}, \href
  {http://adsabs.harvard.edu/abs/2016A%26A...590A..35D} {590, A35}

\bibitem[\protect\citeauthoryear{{van den Bosch}}{{van den
  Bosch}}{2002}]{van-den-bosch2002}
{van den Bosch} F.~C.,  2002, \mn@doi [\mnras]
  {10.1046/j.1365-8711.2002.05171.x}, \href
  {http://adsabs.harvard.edu/abs/2002MNRAS.331...98V} {331, 98}

\makeatother
\end{thebibliography}
\label{lastpage}
\end{document}